\documentclass[reprint,
superscriptaddress,
amsmath,amssymb,
aps, prx,showkeys, showpacs,twocolumn,
]{revtex4-2}

\usepackage{graphicx}% Include figure files
\usepackage{dcolumn}% Align table columns on decimal point
\usepackage{bm,bbm}% bold math
\usepackage{hyperref}% add hypertext capabilities
\usepackage{color}
\usepackage{flushend}

\DeclareMathOperator{\dist}{dist}
\DeclareMathOperator{\mathspan}{span}
\DeclareMathOperator{\argmax}{arg\,max}
\DeclareMathOperator{\rank}{rank}
\DeclareMathOperator{\diag}{diag}

\DeclareMathOperator{\col}{col}
\DeclareMathOperator{\argmin}{argmin}
\DeclareMathOperator{\Gr}{Gr}
\DeclareMathOperator{\cond}{cond}

\newcommand{\dd}{\textnormal{d}}

\newif\ifeditmode

\begin{document}

\editmodetrue
\ifeditmode
  \definecolor{editcolor}{RGB}{0, 0, 255}
\else
  \definecolor{editcolor}{RGB}{0, 0, 0}
\fi

\title{Functional classification of metabolic networks}

\author{Jorge Reyes}
\affiliation{%
Program in Computational and Systems Biology, 
Massachusetts Institute of Technology, Cambridge, MA 02139 
}
\affiliation{%
Department of Mathematics, 
Massachusetts Institute of Technology, Cambridge, MA 02139 
}
\author{J\"orn Dunkel}
\email{dunkel@mit.edu}
\affiliation{%
Department of Mathematics, 
Massachusetts Institute of Technology, Cambridge, MA 02139 
}
\date{\today}

\begin{abstract}
Chemical reaction networks underpin biological and physical phenomena across scales, from microbial interactions to planetary atmosphere dynamics. Bacterial communities exhibit complex competitive interactions for resources, human organs and tissues demonstrate specialized biochemical functions, and planetary atmospheres can display diverse organic and inorganic chemical processes. Despite their complexity, comparing these networks methodically remains a challenge due to the vast degrees of freedom shaped, respectively, by evolution, cell differentiation, and planetary formation. In biological systems, comparative genomics has been pivotal in tracing evolutionary trajectories and classifying organisms via DNA sequences. However, purely genomic classifications often fail to capture functional roles within ecological systems. Metabolic changes driven by nutrient availability highlight the need for classification schemes that integrate metabolic information. Here we introduce a computational framework for comparing chemical reaction networks by measuring distances between the nullspaces of their stoichiometric matrices, corresponding to the spaces of unconstrained steady-state fluxes and conservation laws. Applying this framework to human gut microbiome data shows that nullspace geometry captures organization that is not reducible to phylogenetic relatedness or network-overlap. External phenotype prediction shows that stoichiometric nullspace geometry provides phenotype-specific information complementary to competing descriptors, with phenotype-specific advantages that persist when complete genera are withheld. The generalizability of metabolic Grassmann distances is illustrated by application to chemical reaction networks in human tissue and planetary atmospheres, highlighting its potential for developing functional comparisons from geometric descriptors across diverse chemical reaction systems.
\end{abstract}

\keywords{chemical reaction networks, metabolism, bacterial communities, planetary atmospheres}
\maketitle

\section{Introduction}
 Complex chemical reaction networks are central to the function of living and non-living systems across a wide range of length scales, from microscopic organisms~\cite{McCloskey2013, DalBello2021,Capovilla2023} and tissues~\cite{Jang2019,Thiele2020} to ecosystems~\cite{Feinberg1974,Bonachela2011,Veloz2021,Goyal2023} and planetary atmospheres~\cite{Allen1981,Sole2004,Wong2023}. Recent advances in experimental and computational methods have enabled the comprehensive reconstruction of metabolic processes in various biological systems~\cite{Thiele2010, Chen2021, Heinken2021, Qiang2024}. In bacterial communities, spatiotemporal pyruvate cross-feeding by swarming \textit{Bacillus subtilis} has been observed; bacteria in the swarm front consume their preferred carbon source and deposit pyruvate that bacteria consume in the bulk~\cite{Jeckel2023}. In mice and humans, models of metabolic processes have resolved metabolic cycles and energy use~\cite{Thiele2020,Xiaoxuan2022,Yuan2025}. On the astrophysical scale, the distinctiveness of Earth's atmosphere, from the atmospheres of other celestial bodies in the Solar System, has suggested the development of network-based biosignatures~\cite{Wong2023}. The James Webb Space Telescope and other sources have produced high-quality spectroscopic data that will allow for the chemical characterization of exoplanet atmospheres in remarkable detail~\cite{Yang2024, Teague2025}. Across all of these examples, the breadth of chemistries is shaped by processes that are potentially inaccessible to perturbation or measurement, such as evolution, cellular differentiation, and atmospheric development. Moreover, the vast number of underlying degrees of freedom presents a challenge to the formation of methodical and functional comparisons between chemical reaction networks.

In living systems, inferences of metabolic function from taxonomic classifications are inherently difficult; phylogenetically similar organisms may have vastly different metabolic capabilities~\cite{Vebo2010, Arumugam2011, Monk2013, Bauer2015}. In complex organisms, tissues and organs share the same genetic code and are yet capable of diverse metabolic functions~\cite{Jang2019}. Hence, functional roles cannot readily be ascribed to organisms with similar genetic and evolutionary backgrounds. To tackle this problem, we introduce a conceptual and computational framework for comparing chemical reaction networks, by measuring distances between the nullspaces of their stoichiometric matrices~\cite{Strang1993, StrangLinearAlgebra, Schuster1994, Schilling20002, Polettini2014,Ye2016}, which encode the unconstrained steady-state network fluxes and fundamental conservation laws.

\par
\begin{figure*}[bt!]
\centering
\includegraphics[width=17.8cm]{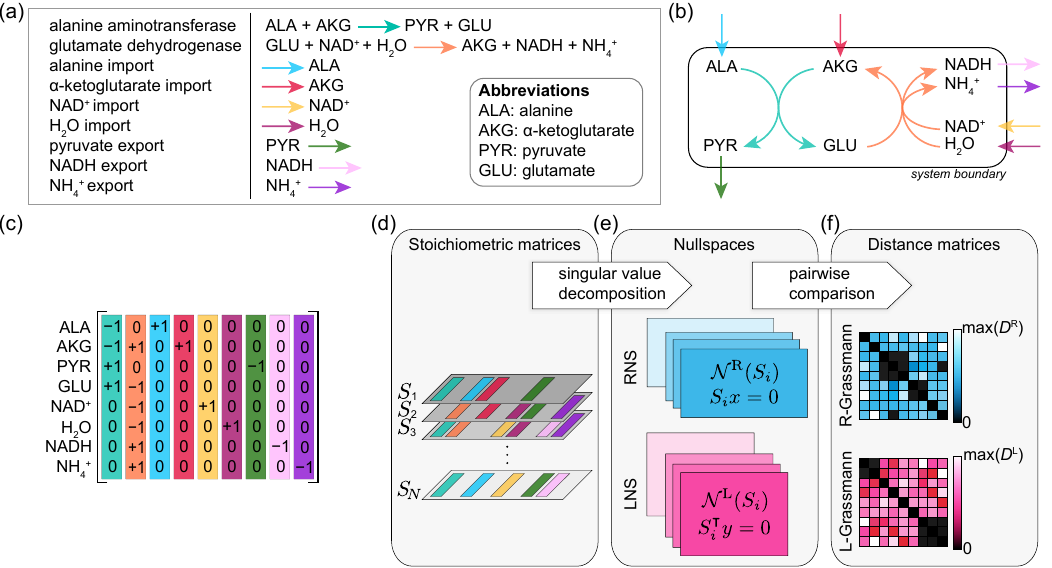}
    \caption{\textbf{Metabolic Grassmann distances are calculated by comparing nullspaces of stoichiometric matrices.}
    Lists of chemical reactions and transport processes (a) are collected in graphs (b) where vertices and edges correspond to chemicals and processes, respectively. The tails and heads of an edge carry information about the number of chemicals consumed and produced by the process, accordingly. The graph representation in turn admits a matrix representation (c): the graph incidence or stoichiometric matrix whose entries are these weights up to a sign which captures whether a metabolite is consumed $(-)$ or produced $(+)$. 
    (d) Row and column-sorted stoichiometric matrices are (e) transformed by computation of their right and left nullspaces---omitting rows and columns of full zeros which correspond to network-specific nonexistent metabolites and processes. (f) Networks are compared pairwise by applying the Grassmann distance metric [Eq.~(\ref{eq:GRASSMANN})] to obtain a distance matrix. Abbr: RNS = right nullspace, LNS = left nullspace. 
    }
    \label{fig:network-grassmann}
\end{figure*}

Chemical reaction networks are naturally described by graphs. Chemical species are represented by vertices and physicochemical processes are represented by weighted and directed hyperedges that capture the direction and stoichiometric quantities of each metabolic process [Figs.~\ref{fig:network-grassmann}(a)-\ref{fig:network-grassmann}(c)]~\cite{Schuster1994,Schilling20002}. Graphs of this flavor admit a matrix representation: the weighted incidence or stoichiometric matrix $S$. Physically, these matrices satisfy the mass-action kinetic differential equation: 
\begin{equation}
\frac{\dd c}{\dd t} = S v ,
\label{eq:MASS-ACTION}
\end{equation} 
where $c$ is a vector of concentrations and $v$ is a vector of fluxes. Each $v_i$ is a sum of directed fluxes for each metabolic process $i$: $v_i = v_i^{(+)} - v_i^{(-)}$ where directed fluxes are proportional to the probability of an encounter between reactants (or products)~\cite{Polettini2014}. If the mathematical forms of these fluxes are known, the concentrations of chemicals in the network are readily obtained~\cite{Bi2023}. Here, we treat fluxes as signed variables. Directionality, irreversibility, and flux bounds can be incorporated as constraints. Specifically, processes proceed in the forward direction if $ v_i^{(+)} > v_i^{(-)}$, while for $ v_i^{(+)} < v_i^{(-)}$ the process is reversed. We note that many complex nonlinear systems can be recast as linear systems of the form of Eq.~(\ref{eq:MASS-ACTION}) where the nonlinear details are absorbed into flux-like functions~\cite{RuizGarcia2021}.

The graph representation of chemical reaction networks makes distance metrics on graphs attractive choices. However, not all graph distance metrics are suitable for directed graphs with hyperedges; applications of these metrics typically ignore directionality or stoichiometry~\cite{Tantardini2019, Wills2020}. Other metrics opt for computational tractability, such as reducing the scope to comparisons of the presence/absence of metabolic processes~\cite{Mazumdar2013, Bauer2015} or feature vectors of topological measures derived from a graph-theoretic approach~\cite{Machicao2018,Wong2023}. Flux-based metrics~\cite{Ebenhoh2009, Beguerisse2018, Ramon2023} are an alternative approach for comparing chemical reaction networks, yet these require system-specific optimization criterion to calculate. For metabolic modeling, this typically involves maximizing flux through a biomass or synthetic biochemical process. In imposing this constraint, the richer mathematical structure of the underlying flux space is ignored. We avoid these information losses by leveraging advances in parallel computing and numerical linear algebra to make calculations of stoichiometric nullspaces tractable for large datasets~\cite{Bezanson2017, Reuther2018}. At this point, one may ask why focus on nullspaces of stoichiometric matrices?   

\section{Stoichiometric Nullspaces}

Principally, the stoichiometric nullspaces have clear physical interpretations in terms of the mass-action picture of Eq.~(\ref{eq:MASS-ACTION}). The right nullspace of a stoichiometric matrix satisfies the nonequilibrium unconstrained steady-state flux condition required by flux balance‐based approaches~\cite{Schuster1994,Schilling20002,Motter2008},
\begin{equation*}
    \mathcal{N}^\text{R} (S) = \{ v \, : \, Sv = 0  \}. 
\end{equation*}
Namely, the right nullspace contains linear combinations of processes that result in net zero consumption and production of all chemical species. This is equivalent to currents satisfying Kirchhoff's current law in electrical circuits~\cite{Polettini2014} (Appendix~\ref{sec:stoich-nullspaces-transport-networks}). The right nullspace coordinate representation is sensitive to the choice of column-wise normalization of the stoichiometric matrix: 
\begin{equation}
    S' = S \Lambda ,
    \label{eq:REACTION-RESCALING}
\end{equation}
where $\Lambda = \diag(\lambda_1,\ldots,\lambda_n)$ is a positive definite diagonal matrix whose entries weighs each metabolic process accordingly. This is not the same as a change to the right nullspace, but rather its representation. Nonetheless, any metric defined on the right nullspace is subject to this sensitivity. By convention, the stoichiometries of chemical processes are either reducible by their greatest common divisor or, for electron-balanced reactions, balanced so that electron carriers have unity stoichiometry, potentially resulting in fractional coefficients for other chemicals. Moreover, some processes contain empirically derived non-integer stoichiometries. As such, right nullspaces herewithin are subject to this conventional normalization choice, as reported by each dataset, which is consistently maintained within each dataset considered in this work. Although alternate normalization choices lose the interpretation of minimal integer-many molecules interacting, we consider such normalization choices in Appendix~\ref{sec:stoich-nullspaces-transport-networks}. Under the framework considered in this work, we find that the absolute difference in distances between the right nullspaces of pairs of stoichiometric matrices and their normalized counterparts is upper-bounded by a function of $\Lambda$ which is empirically unsaturated (Appendix~\ref{sec:stoich-nullspaces-transport-networks}).

Metrics derived from flux balance analysis solutions~\cite{Ebenhoh2009, Beguerisse2018, Ramon2023} are a promising initial step in this direction. These metrics require the identification of right nullspace elements that are optimal under prescribed external influxes and a particular optimization criterion. As such, they are powerful descriptors of condition-specific metabolic states. However, FBA-derived metrics compare select flux vectors rather than the full space of unconstrained steady-state fluxes. A minimal chemoton example in Appendix~\ref{sec:chemoton} illustrates this distinction explicitly: distances between FBA solutions depend on nutrient influxes and objective function choice, whereas a distance over the nullspace directly compares the underlying unconstrained steady-state flux spaces without selecting a particular optimum. The latter is particularly important for systems where there is no obvious optimization criterion, such as planetary atmospheres.

Comparisons of right nullspaces do not inherently account for inequality constraints such as irreversibility, thermodynamic feasibility, or condition-specific flux bounds. Instead, they can only be interpreted as a comparison of the unconstrained flux space, rather than the flux polytope used in constraint-based modeling \cite{Schuster1994, Schilling20002}:
\begin{equation*}
\mathrm{Polytope}(S) = \{ v \, : \, Sv = 0 \ \text{and} \ 
v_\text{lb} \leq v \leq v_\text{ub}
\},
\end{equation*}
where $v_\text{lb}$ and $v_\text{ub}$ are flux lower and upper bounds respectively, and the inequality constraint is performed element-wise. Further mathematical developments are required to analyze the differences in such constrained spaces and is outside the scope of this work. Thus, although we cannot address the distribution of such fluxes under a range of influxes, inequality constraints, or the assumption of optimality, the general space of unconstrained steady-state fluxes should not be ignored in developing intuition of potential metabolic strategies.

Contrast this with the left nullspace which consists of conservation laws for pools of chemicals~\cite{Famili2003, Polettini2014, Avanzini2023, Srinivas2024, Srinivas2024PRE},
\begin{equation*}
\mathcal{N}^\text{L}(S) = \{ w \, : \, S^\intercal w = 0 \}.
\end{equation*} 
Indeed, we see that for an element $w$ of the left nullspace: 
\begin{equation*}
    w \cdot \frac{\dd c}{\dd t} = w \cdot (Sv) = (S^\intercal w) \cdot v  = 0 \implies w \cdot c = \text{constant}.
\end{equation*}
Previous characterizations of the left nullspace associate it with biological properties such as energy and redox potential, which are essential to meet energetic demands and are maintained by specific pools of molecules~\cite{Maloney1974, Famili2003, Glasser2017, Horak2024}. Unlike the right nullspace, the left nullspace is invariant under the rescaling of Eq.~(\ref{eq:REACTION-RESCALING}). 

For the left nullspace, we must also consider network closure: should we include exchange processes across the system boundary [Fig.~\ref{fig:network-grassmann}(b)]? In an open network, the system is forced away from thermodynamic equilibrium by the exchange of mass and energy with the environment. As such, exchange processes can break conservation laws that would otherwise be present in the corresponding closed system~\cite{Famili2003, Polettini2014}. Consequently, the open-network left nullspace may be low-dimensional or even trivial, not because the internal network lacks structure, but because exchange fluxes allow for net accumulation or depletion of species pools. This exchange of resources with the environment confers the system with a dynamical flexibility that allows the chemical species to, in principle, move through the full concentration space. Because our aim is to compare the right and left nullspaces of the same physical system, we use the open network as the default representation for both nullspaces unless stated otherwise. Under this convention, the right nullspace describes unconstrained steady-state flux combinations and the left nullspace describes conservation relations that remain valid in the presence of environmental exchange. 

In some settings, the open-network left nullspace is trivial (Appendix~\ref{sec:stoich-nullspaces-transport-networks}), reducing its discriminatory power. To examine conservation structure masked by environmental exchange, we therefore include, as a secondary analysis, the left nullspace of the corresponding closed network obtained by removing exchange processes. Nonetheless, the closed-network left nullspace is not the conservation law structure of the physically open biological system, rather it is a descriptor of latent internal conservation structure in the absence of environmental exchange. All main analyses and claims reported here use open networks; closed-network results are included as an auxiliary analysis of latent internal conservation structure. 

We further discuss the stoichiometric nullspaces of transport networks, including electrical circuits and mechanical networks in Appendix~\ref{sec:stoich-nullspaces-transport-networks}. The physical perspective offered by both nullspaces suggests that they are suitable candidates for developing a functional classification scheme of chemical reaction networks. 

\section{The Grassmann Distance}

Distances between linear subspaces have already appeared in analyses of electroencephalogram signals~\cite{Figueiredo2010}, network security~\cite{Sharafuddin2010} and undulatory worm locomotion~\cite{Cohen2023}. The classic method of comparison uses that linear subspaces of dimension $k$ embedded in $\mathbb{R}^n$ are elements of the Grassmannian manifold $\Gr(k,n)$ with a geodesic metric that is computed by singular value decomposition. This Grassmann distance metric generalizes in a nontrivial manner to linear subspaces of all dimensions, regardless of the value of $k$ and $n$. The generalized Grassmann distance is defined on the manifold of linear subspaces of all dimensions, elements of the doubly infinite Grassmannian $\Gr(\infty,\infty)$. On $\Gr(\infty,\infty)$, the geodesic distance between the $k$-dimensional subspace with orthonormal basis matrix $A\in \Gr(k,\infty)$ and the $\ell$-dimensional subspace with orthonormal basis matrix $B\in \Gr(\ell,\infty)$ is~\cite{Ye2016}: 
\begin{equation} \textstyle
d_{\Gr(\infty,\infty)}(A,B) =  
\sqrt{\left| k-\ell \right| \pi^2/4 + \sum_{i = 1}^{\min(k,\ell)} \theta_i^2 }.
\label{eq:GRASSMANN}
\end{equation}
Here the $\theta_i$ are the principal angles between the subspaces and are obtained from: 
\begin{gather*}\textstyle
    \Theta = \diag(\theta_1,\ldots ,\theta_{\min (k,\ell)})  = \cos^{-1} \Sigma  \\
    \text{where} \ \ \ A^\intercal B = U \Sigma V^\intercal .
\end{gather*}
If we consider $k\leq \ell$, the Grassmann distance metric provides a distance from the $\ell$-dimensional subspace $B$ to the furthest $\ell$-dimensional subspace that contains the $k$-dimensional subspace $A$. The symmetric statement is also true: it is the distance from the $k$-dimensional subspace $A$ to the furthest $k$-dimensional subspace contained in the $\ell$-dimensional subspace $B$~\cite{Ye2016}. 

Fundamentally, the Grassmann distance [Eq.~(\ref{eq:GRASSMANN})] is comprised of a dimension gap, the difference in the dimension of the subspaces, and an angular term that corresponds to correlations between the basis vectors, the sum of the squared principal angles. From the rank-nullity theorem, we know that the dimensions of the stoichiometric nullspaces are partially determined by the number of chemical species and metabolic processes (Appendix~\ref{sec:dist-components}). In this manner, the Grassmann distance on stoichiometric nullspaces may capture size differences between networks. In general, two networks have zero Grassmann distance only when the corresponding nullspaces are identical. The distance increases either because the nullspaces have different dimensions or because basis directions in the common-dimensional component rotate away from one another. Herewithin, we will refer to the geodesic distance metric between right and left stoichiometric nullspaces, in the doubly infinite Grassmannian, as R-Grassmann and L-Grassmann, respectively [Figs.~\ref{fig:network-grassmann}(d)-\ref{fig:network-grassmann}(f)].

As alluded to, we show in Appendix~\ref{sec:stoich-nullspaces-transport-networks} that rescaling stoichiometric matrices according to Eq.~(\ref{eq:REACTION-RESCALING}) results in changes to the R-Grassmann distance that are bounded by the entries of the rescaling matrices. Alternative distance metrics over the doubly-infinite Grassmannian manifold exist \cite{Ye2016}, however all such distance metrics on right nullspaces are sensitive to rescalings of the form of Eq.~(\ref{eq:REACTION-RESCALING}). Additional mathematical tools are needed to construct metrics invariant to these rescalings, although tropicalization of the Grassmannian manifold is a promising candidate \cite{Speyer2004,Bhatt2025}. 

\section{Grassmann Distances on Chemical Reaction Networks}

We now move towards applications of the Grassmann distance to chemical reaction networks. First, we examine the effect of \textit{in silico} genetic perturbations in the genome of the model organism \textit{Escherichia coli} K-12 MG1655 on these metabolic Grassmann distances~\cite{Heinken2023}. We then consider genetically diverse organisms present in the AGORA2 (assembly of gut organisms through reconstruction and analysis, version 2) dataset which serves as a metabolic knowledge base for the human gut microbiome~\cite{Heinken2023}. We show that the resulting Grassmann distances on nullspaces can differ substantially from taxonomic structures obtained from comparisons of the genetic background. To establish the potential of Grassmann distances for developing functional classifications, we identify metabolic processes that annotate clusters and we examine their predictive power on external binary phenotype annotations, as well as for the computationally tractable Jaccard distance~\cite{Mazumdar2013, Bauer2015}. A theoretical treatment of Ganti's chemoton~\cite{Ganti1975,Csendes1984} as a toy model (Appendix \ref{sec:chemoton}) explicitly demonstrates that set-overlap metrics can be insensitive to stoichiometric differences between networks with distinct metabolic strategies, whereas Grassmann distances detect these differences through changes in nullspace geometry. The final application of the Grassmann distance on the chemical reaction networks of human tissues~\cite{Thiele2020} and planetary atmospheres~\cite{Wong2023} serves as a proof of concept and highlights the applicability of Grassmann distances to systems of different length scales and fields. 

The bacterial networks considered here were obtained from the AGORA2 dataset~\cite{Heinken2023}. We used the 635 out of the 7302 published human-associated metabolic networks~\cite{Heinken2023} which have complete comparative genomics as described in Refs.~\cite{Heinken2021} and~\cite{Heinken2023}. We refer to these networks as mini-AGORA2 (Supplemental Table 1). Of these, 619 networks were unique and are used for all following analyses unless otherwise stated. The human tissue networks presented in this work were obtained from the Harvey and Harvetta reconstructions of human metabolism~\cite{Thiele2020}. The stoichiometries of the biological metabolic processes are unweighted [Eq.~(\ref{eq:REACTION-RESCALING})] and taken as reported in Refs.~\cite{Heinken2023} and~\cite{Thiele2020}. The planetary atmosphere networks were derived from~\cite{Wong2023} where chemicals without molecular formulas were removed and catalytic chemicals that appeared as both reactant and product were reduced to simplify stoichiometry. Any resulting processes without either reactants or products were taken to be exchange processes with the environment. 

 Without proper conditioning, any chemical reaction network may not be physically admissible. We ensure all networks considered here satisfy flux and stoichiometric consistency using methods available in the \textsc{COBRA Toolbox}~\cite{Acuna2009, Vlassis2014, Thiele2014, Heirendt2019, Fleming2022} with \textsc{Gurobi v12.0.3}~\cite{gurobi}. Flux consistency requires that each process of a chemical reaction network is active, realizing nonzero flux, in at least one flux distribution~\cite{Acuna2009, Vlassis2014, Fleming2022}. We remove flux inconsistent internal (non-exchange) processes to ensure we only consider active metabolic processes~\cite{Vlassis2014, Thiele2014}. Similarly, reaction databases may contain stoichiometric inconsistencies~\cite{Gevorgyan2008}, where the stoichiometry of processes is inconsistent with mass conservation. To that end, we identify and correct, or remove, stoichiometric inconsistent and elementally imbalanced internal processes~\cite{Thiele2014}. For each type of network, we placed rows and columns of stoichiometric matrices alphabetically by chemical species and metabolic process name, respectively, according to a globally ordered coordinate list using I/O functions implemented in the \textsc{COBRA Toolbox v3.5} (commit \texttt{f301a51eaad06b141e7357fead237560a2dda7cf}) for \textsc{Matlab 2024a} (MathWorks)~\cite{Heirendt2019}. For bookkeeping purposes only, any chemical and process not present in the network introduces a row or column of zeros accordingly; these additional rows and columns are otherwise not considered in downstream network analyses. Biomass reactions are treated as the same process for human gut microbiome networks, even if the stoichiometries differ. All information on biological metabolic processes and chemicals were obtained from the Virtual Metabolic Human (VMH) database and is presented here in its nomenclature~\cite{Heinken2021}.

\begin{figure*}[bt!]
\centering
\includegraphics[width=17.8cm]{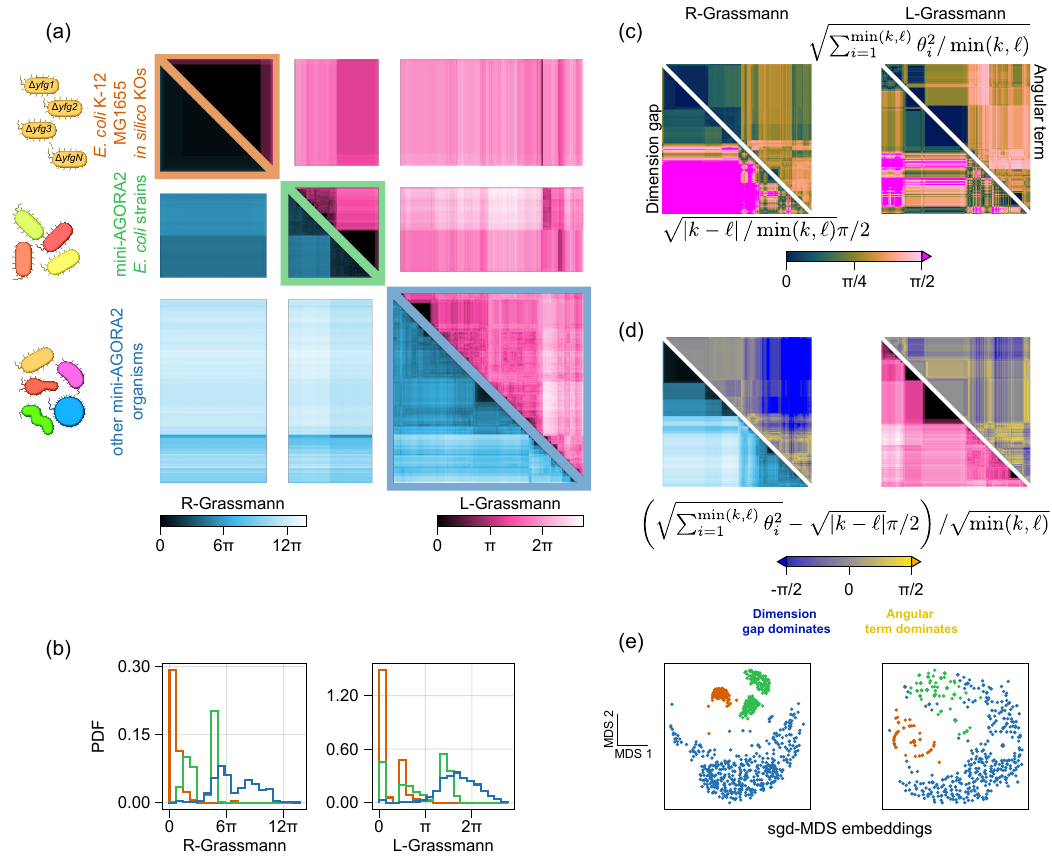}
\caption{\textbf{Metabolic Grassmann clusters are the result of competition between the dimension gap and angular term of the distance metric which is robust to genetic knockout perturbation.} (a) Metabolic Grassmann distance matrices are computed for organisms at different scales of genetic similarity: computationally viable \textit{Escherichia coli} K-12 MG1655 \textit{in silico} KOs (orange, top), \textit{E. coli} strains in mini-AGORA2 (green, center), and all other mini-AGORA2 organisms (blue, bottom). (b) Distance distributions show preferences for larger distances with increasing genetic diversity. Each block diagonal matrix is sorted according to a hierarchical clustering with average linkage and sets the corresponding sorting for the off diagonal block matrices. (c) Squared Grassmann distances shown in (a) are decomposed as a dimension gap (left) and angular term (right) for both nullspaces. Matrices are sorted according to a matrix hierarchical clustering with average linkage. (d) A component contribution score, shown in Eq. (\ref{eq:COMPONENT-DOMINANCE}), computed from (c) highlights that the dimension gap and angular term dominates differently across network comparisons. Distances that cannot be dominated by the angular term due to the structure of the Grassmann metric are shown in magenta. (e) Stochastic-gradient descent multidimensional (sgd-MDS) embeddings are shown for all organisms considered in (a). Bacteria images were obtained and modified from Ref.~\cite{LeMercier2022} under a Creative Commons Attribution 4.0 International License. 
} \label{fig:distancemats}
\end{figure*} 

To calculate the metabolic Grassmann distances, we compute basis vectors for the right nullspace using an SVD-based nullspace function derived from the \textsc{LinearAlgebra.jl} package in \textsc{Julia}~\cite{Bezanson2017} (Appendix~\ref{sec:num-stability}), omitting columns and rows of zeros which were introduced to account for nonexistent processes and metabolites, respectively. Nullspaces are computed in this reduced coordinate system. To keep all basis vectors across networks in a shared ambient space, we insert zeros in indices that correspond to those nonexistent processes and metabolites. This means that all Grassmann comparisons are made in a shared ambient coordinate space; without this alignment, comparing principal angles would be meaningless. Note that this is explicitly not equivalent to computing nullspaces from a globally padded matrix. Doing so would introduce additional nullspace directions. The left nullspace of the matrix is obtained by calculating the right nullspace of the transposed matrix $M^\intercal$. For the Jaccard distance, we look at lists of metabolic processes $S$ and $T$ and compute: 
\begin{equation*} \textstyle
    \dist_\text{Jaccard}(S,T) = 1 - {\#(S \cap T) }/{\#(S \cup T)}.
\end{equation*}
Here $\#$ is the set cardinality function which counts the number of elements in the set. 

\begin{figure*}[bt!]
\centering
\includegraphics[width=17.8cm]{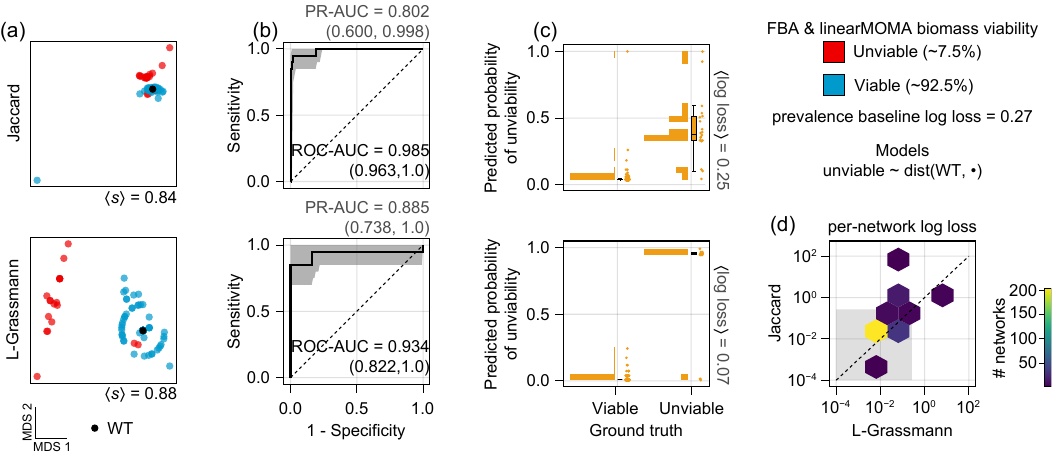}
    \caption{\textbf{\textit{In silico Escherichia coli} KO viability is readily identified by the L-Grassmann distance.} Viability  of \textit{E. coli} KOs are assessed by flux balance analysis (FBA) and minimization of metabolic adjustment (MOMA). A network is viable if it realizes a nonzero biomass flux in any flux distribution. We observe no differences in viability arising from the choice of FBA versus MOMA. Duplicate metabolic networks (networks with identical metabolic processes) are excluded from the following analysis. (a) Mean silhouette scores $\langle s \rangle$ of the \textit{E. coli} distance matrix, with viability as cluster assignment, reveals that the L-Grassmann distance best captures the geometric separation in KO viability. Silhouette scores are found to be statistically significant ($p < 0.01$) by permutation significance testing with finite-sample correction. (b) ROC curves and ROC-AUC quantify the ability of the metabolic distances to wild-type (WT) to discriminate between viable and nonviable \textit{E. coli} networks where (c) predicted probabilities were obtained by leave-one-out cross-validation with Firth logistic regression. Note that the Jaccard distance to WT and the number of metabolic processes present, a simpler predictor, are related by a linear transformation. ROC-AUC from both predictors are found to not be significantly different ($p> 0.05$) by a paired bootstrap test with finite-sample correction. Due to the class imbalance, we also compute PR-AUCs which are also not found to be statistically different by a paired bootstrap significance test with finite-sample correction. AUC 95\% confidence intervals are reported in parentheses. Mean log loss was used to assess probabilistic performance where lower values indicate better performance by the L-Grassmann distance.  Box-and-whisker plots show the quartiles with error bars corresponding to 1.5$\times$IQR beyond Q1 and Q3. (d) The per-network log-loss density plot compares Jaccard and L-Grassmann distances directly, indicating that the majority of networks realize lower log loss with the L-Grassmann distance. The gray box highlights per-network log losses less than the prevalence baseline.}
    \label{fig:ecoli-ko-viability}
\end{figure*}

\section{Robustness of metabolic Grassmann distances to genetic variations\label{sec:GENETIC-VARIATION}} 

To illustrate the robustness of the Grassmann distances to genetic perturbations in the form of gene knockouts (KOs), we use the AGORA2 metabolic network of the model organism \textit{E. coli} K-12 MG1655. Genome-scale metabolic reconstructions, such as those in AGORA2, contain gene-protein-reaction (GPR) rules that map the presence of genes to the presence of metabolic processes ~\cite{Heinken2023}. An \textit{in silico} single gene knockout is obtained by evaluating these GPR rules with all other genes present and identifying which metabolic processes are as a result absent in the knockout. Of the 747 single-gene KOs, 242 had no change in metabolic processes under the evaluation of GPR rules, whereas 505 were altered. After removing duplicate networks, 267 unique KO networks remained. We assess network viability with flux balance analysis (FBA) and minimization of metabolic adjustment (MOMA) using methods available in the \textsc{COBRA Toolbox}~\cite{Heirendt2019} with \textsc{Gurobi v12.0.3}~\cite{gurobi}. A network is deemed to be viable \textit{in silico} if it realizes a nonzero biomass flux in a flux distribution obtained from FBA or MOMA. Among the 267 unique KO networks, 247 ($\approx$93\%) were computationally viable; they admit nonzero biomass flux under AGORA2 lower and upper bound fluxes. Among these, no network had R-Grassmann distance zero to WT and 201 had L-Grassmann distance zero to WT, indicating that most \textit{E. coli} GPR perturbations leave the stoichiometric left nullspace unchanged. The latter is not a limitation of the L-Grassmann distance but rather an indication of the robustness of the conservation law structure, represented by stoichiometric left nullspaces, to such perturbations. This makes the L-Grassmann distance a promising candidate for discerning the viability of organisms. Furthermore, two-dimensional embeddings of the metabolic distance matrices, via stochastic gradient descent multidimensional scaling (sgd-MDS, Appendix~\ref{sec:sgd-MDS}) and mean silhouette scores using viability as cluster assignment [Fig.~\ref{fig:ecoli-ko-viability}(a)], further suggest that the L-Grassmann and Jaccard distance considered here may be capable of distinguishing computationally viable and unviable networks. 

Overall, we observe that most computationally viable single gene KOs lead to no or minimal changes in nullspaces [Fig.~\ref{fig:distancemats}(a)-\ref{fig:distancemats}(b)] relative to the differences observed in non-\textit{E. coli} networks. The exception here corresponds to the right nullspace of $\Delta$\textit{uidA} that otherwise encodes for $\beta$-glucuronidase which, in the context of the human gut microbiome, modifies hydrophilic molecules for elimination by the host~\cite{Blanco1985, Yang2017,Gao2022}. Previous work indicates that the \textit{uidA} gene is nearly ubiquitous in \textit{E. coli} isolates from treated and raw water sources~\cite{Martins1993}. Mechanistically, of the single gene KOs, the $\Delta$\textit{uidA} network has the largest effect size in the number of metabolic processes; almost 10$\times$ greater than the next largest effect size. Together, this is consistent with its R-Grassmann distance from other \textit{E. coli} KOs. We further validate that the \textit{E. coli} KO Grassmann distances are numerically stable under changes in singular-value thresholding and nullspace rank selection is assessed in Appendix D [Fig.~\ref{fig:sensitivity}].

To assess the viability distinguishing power of the L-Grassmann and Jaccard distances, we examine the prediction of \textit{in silico} viability in a supervised leave-one-out cross-validation (LOOCV) framework [Fig.~\ref{fig:ecoli-ko-viability}] using \textsc{R} libraries \textsc{logistf}, \textsc{pROC}, and \textsc{precrec}. Within each LOOCV fold, we take the score to be the distance from the wild-type (WT) network:
\begin{equation*}
   \operatorname{score}(\mathcal{N}_i) = \dist(\mathcal{N}_i, \mathcal{N}_\text{WT})
\end{equation*}
Score-standardization and Firth regression were computed exclusively from the within-fold training set. Folds in which the training set did not contain both viability classes were set to the prevalence baseline of the in-fold samples. Out-of-fold probabilities were obtained from Firth logistic regression models fitted using only the corresponding outer training data. Predictor scores were standardized using training-fold statistics before model fitting, and the same transformations were applied to the held-out network. The predictive performance was then calculated from the pooled out-of-fold probabilities, such that every eligible network contributed exactly one probability under each predictor. We evaluated each predictor separately using models of the form
\begin{equation}
    \operatorname{logit}P(Y=1)
=
\beta_0+\beta_A z_A, 
\label{eq:SIMPLE-LOGISTIC-PREDICTOR}
\end{equation} 
where $z_A$ is the training-fold-standardized score derived from the L-Grassmann or Jaccard predictor and $Y$ is the binary viability state. The less prevalent class of unviable networks was fixed as the positive class, $Y=1$. 

We evaluated the discrimination, using ROC-AUC and PR-AUC [Fig.~\ref{fig:ecoli-ko-viability}(b)], and probabilistic accuracy, using log loss [Fig.~\ref{fig:ecoli-ko-viability}(c)]. Receiver operating characteristic (ROC) curves summarize the ability of each predictor to discriminate viable from nonviable \textit{E. coli} networks, with the area under the ROC curve (ROC-AUC) quantifying ranking performance. Log loss is used to assess probabilistic performance, with lower values indicating better calibrated and more confident predictions. ROC-AUC is insensitive to the positive-class prevalence, but can provide an optimistic summary under substantial class imbalance. We therefore also report the area under the corresponding precision-recall (PR) curves, whose no-skill baseline is the positive-class prevalence. We show the per-network differences in log loss contribution between models [Fig.~\ref{fig:ecoli-ko-viability}(d)] and identify that nearly all per-network log loss values of the Jaccard predictor exceed their corresponding L-Grassmann values. Note that the Jaccard distance score is related to the number of metabolic processes in the network by a simple linear transformation; thus the Jaccard comparison also serves as a process-count baseline:
\begin{align}
    \dist_\text{Jaccard}(S_\text{WT}, S_\text{KO}) &= 1 - \frac{\#(S_\text{WT}\cap S_\text{KO})}{\#(S_\text{WT}\cup S_\text{KO})} \notag \\
    &= 1 - \frac{\#(S_\text{KO})}{\#(S_\text{WT})} 
    .
    \label{eq:KO-JACCARD}
\end{align}
Statistical significance of the ROC-AUC and PR-AUC differences was evaluated by nonparametric bootstrap testing of the out-of-fold predictions. Within each bootstrap replicate, the same sampled observations and outcomes were used for both models in a comparison, and the difference in the relevant performance statistic was recomputed. Bootstrap $p$-values were calculated with finite-sample correction. 

We find that while the L-Grassmann and Jaccard distances do not show statistically significant differences in the discrimination of viability [Fig.~\ref{fig:ecoli-ko-viability}(b)], both distance provide improvements on the prevalence baseline and the L-Grassmann distance has a considerably improved  probabilistic performance [Fig.~\ref{fig:ecoli-ko-viability}(c)-\ref{fig:ecoli-ko-viability}(d)]. As stoichiometric nullspaces are inherently dependent on network connectivity, our results are consistent with previous work that has shown that metabolic network topology is predictive of knockout viability \cite{Wunderlich2006}.

To go beyond small genetic differences, let us focus on mini-AGORA2, a subset of the metabolic networks of AGORA2, which represents 635 genetically distinct microbes from the human gut microbiome~\cite{Heinken2023}. We cross-validate that each microbe is associated with humans using the Joint Genome Institute (JGI) Integrated Microbial Genomes \& Microbiomes (IMG/M) database~\cite{Chen2023}. With mini-AGORA2, we examine microbial metabolism as shaped by evolution and the human gut environment. We find that compared to the computationally viable KOs of \textit{E. coli} K-12 MG1655, the mini-AGORA2 \textit{E. coli} strains realize larger distances in both nullspaces, which are further augmented when considering all mini-AGORA2 networks [Fig.~\ref{fig:distancemats}(b)]. Two-dimensional embeddings of the combined Grassmann distance matrix reveal that networks of viable \textit{E. coli} KOs and mini-AGORA2 \textit{E. coli} strains cluster away from other mini-AGORA2 networks [Fig.~\ref{fig:distancemats}(e)]. This suggests that high genetic similarity may be sufficient to produce similar nullspaces. 

To investigate the mathematical origin of clusters, we consider Grassmann distances to be the modulus of a complex number $f = f_\theta + i f_\text{dim}$, where the real component $f_\theta$ and imaginary component $f_\text{dim}$ are the angular term and dimension gap of Eq.~(\ref{eq:GRASSMANN}), respectively. Scaling the components by the number of principal angles, we obtain
\begin{align*}
\textstyle 
    \hat{f}_\text{dim} &= \textstyle \sqrt{ |k-\ell|/ \min(k,\ell) } \, \pi/2 \ \ \ \text{and} \\
    \hat{f}_\theta &= \textstyle \sqrt{\sum_{i=1}^{\min{(k,\ell) }}\theta_i^2 / \min(k,\ell)},
\end{align*}
which are shown for the combined matrix of Fig.~\ref{fig:distancemats} in Fig.~\ref{fig:distancemats}(c). The first balances the dimension gap against the number of principal angles, whereas the second is equivalent to a root-mean-squared principal angle. Notably, when $\hat{f}_\text{dim} > \pi/2$ the distance is structurally incapable of being dominated by the angular term [Fig.~\ref{fig:distancemats}(c)]. Importantly, these quantities require $\min(k,\ell) > 0$, which is satisfied by every pairwise bacterial metabolic distance in our analysis. 

We can define a Grassmann component dominance descriptor 
\begin{equation}
   \Delta f = \hat{f}_\theta - \hat{f}_\text{dim}
   \label{eq:COMPONENT-DOMINANCE}
\end{equation}
which reports the difference in each component averaged over the number of principal angles. If $\Delta f >0$, the angular term dominates and if $\Delta f < 0$, the dimension gap dominates. In the case of nullspace equality, both components are zero. We find that each component dominates in different regions of the distance matrices [Fig.~\ref{fig:distancemats}(d)]. Additionally, regions of high similarity, but not equality, correspond to either the angular term dominating in contribution or a balancing of components, whereas regions of low similarity correspond to dimension gap dominating. Moreover, from a comparative analysis of the contribution to the Grassmann distances within- and between-clusters, we find that between-cluster distances are dominated by the dimension gap contribution and this dominance is lost when examining within-cluster distances (Appendix~\ref{sec:dist-components}). Ultimately, clusters are due to these competing contributions to the Grassmann distance, rather than being solely a consequence of network size and nullity.

\section{Inequivalence of genetic and metabolic distances}

\begin{figure*}[bt!]
\centering
\includegraphics[width=17.8cm]{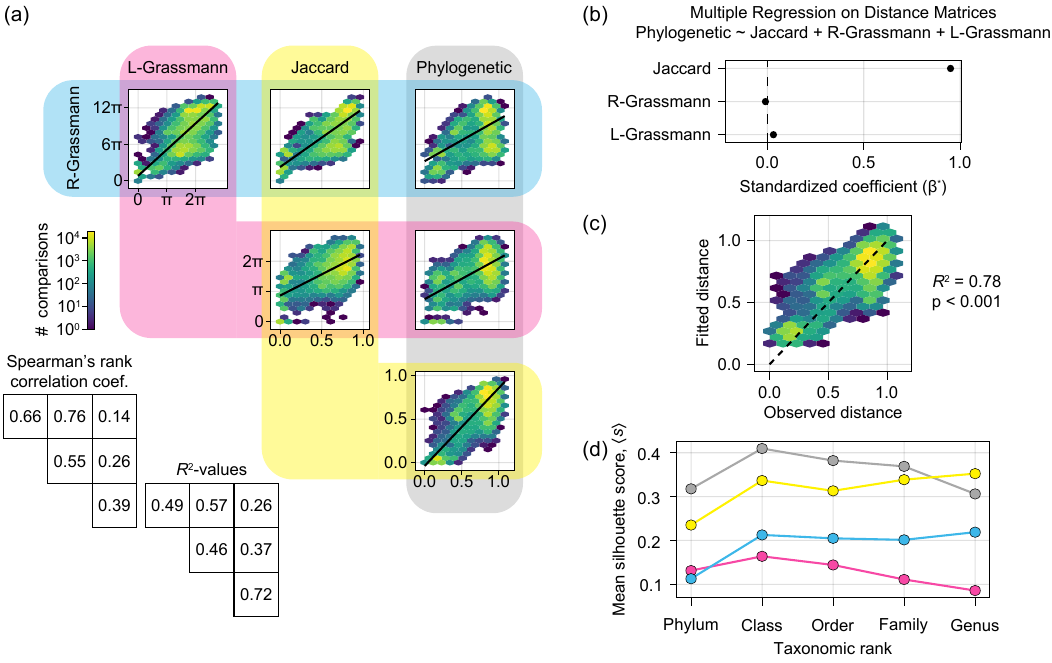}
    \caption{\textbf{Inequivalence of metabolic and phylogenetic metrics in mini-AGORA2 organisms.} (a) The Jaccard distance linearly correlates the most and among the metabolic distances displays the largest rank correlation to the phylogenetic distance which suggests that it is not the best choice for substantially distinguishing organisms beyond genetic differences. The line of best fit is shown in red with corresponding $R^2$ values and Spearman's rank correlation coefficients on the bottom left. 
(b) In the multiple regression on distance matrices (MRM) model $\text{Phylogenetic} \sim \text{R-Grassmann} + \text{L-Grassmann} + \text{Jaccard}$, the predictors jointly explained 78\% of the variation in phylogenetic distance ($R^2 = 0.78$, permutation $p < 0.001$). Only the Jaccard distance showed a significant conditional effect ($\beta^* = 0.95$). (c) Observed and fitted phylogenetic distances from the MRM model are compared where each point represents one pairwise comparison between metabolic networks and the dashed line indicates the identity relationship. (d) Mean silhouette scores $\langle s \rangle$ are computed for all distances using taxa at each taxonomic rank as cluster assignments. Phylogenetic distance outperforms both Grassmann distances across all taxonomic ranks. 
    Here we compare all 635 organisms of mini-AGORA2, since phylogenetic distance is not guaranteed to be zero for organisms with identical networks.
    }
    \label{fig:compare-metrics}
\end{figure*}

To investigate the differences between Grassmann, Jaccard and genetic distances, we perform linear regression analyses with mini-AGORA2 distances for all pairs of metrics. To compute genetic distance, we first infer phylogenies from genome sequences. We take the approach used in Ref.~\cite{Bauer2015} where the PhyloPhlAn pipeline~\cite{Asnicar2020} is applied to the available genome sequences as well as the genome sequence of the archaebacteria \textit{Methanobrevibacter smithii} ATCC 35061. We root the resulting tree using \textit{M. smithii} as the outgroup using \textsc{PhyloNetworks.jl}~\cite{SolisLemus2017} and compute pairwise tree distances. We then use the square root of these tree distances as it is Euclidean-like~\cite{deVienne2011} and provably metric (Appendix~\ref{sec:square-root-metric}). Here we use pairwise distances between all 635 networks of mini-AGORA2. 
 
We verify that small phylogenetic differences between two organisms can produce appreciable differences in nullspaces [Fig.~\ref{fig:compare-metrics}(a)]. Moreover, we find that the Grassmann distances display less linear phylogenetic predictive power than the Jaccard distance [Fig.~\ref{fig:compare-metrics}(a)]. The latter is consistent with existing work that has shown an exponential relationship between the Jaccard distance and the cophenetic distance in human gut microbiome metabolic networks~\cite{Bauer2015}. The cophenetic tree is an alternative genetic distance derived from phylogenetic trees that uses the height of the most recent common ancestor~\cite{Zaneveld2010}. Spearman's rank correlation coefficient is consistent with the exponential relationship described in Ref.~\cite{Bauer2015} and further shows stronger phylogenetic correlation between the Jaccard distance than with the Grassmann distance [Fig.~\ref{fig:compare-metrics}(a)]. In a multiple regression on distance matrices (MRM) analysis [Figs.~\ref{fig:compare-metrics}(b)-\ref{fig:compare-metrics}(c)] performed using the \textsc{R ecodist} library, the metabolic distances are put in direct competition with each other. While the metabolic distances jointly explain 78\% of the observed variance in phylogenetic distance, we find that only the Jaccard distance shows a significant conditional effect on predicting phylogenetic distances.

We then compare each distance by assigning organisms to different clusters according to their taxa at different phylogenetic taxonomic ranks: from phylum to genus. To validate goodness of clustering on the basis of phylogeny, we compute mean silhouette scores using the \textsc{Julia Clustering.jl} package~\cite{Rousseeuw1987, Bezanson2017}. We find that the Grassmann metabolic distances produce values smaller than the phylogenetic distance, indicating a loss of adherence to these categories that is not entirely observed with the Jaccard distance [Fig.~\ref{fig:compare-metrics}(d)]. 

Together, these results show that the Grassmann and Jaccard distances capture different aspects of network variation. In particular, the Jaccard distance remains more closely aligned with phylogenetic relatedness, whereas the Grassmann distances retain variation less directly explained by phylogeny. Whether that additional variation is functionally informative must be assessed using external phenotype data.

\begin{figure*}[bt!]
\centering
\includegraphics[width=17.8cm]{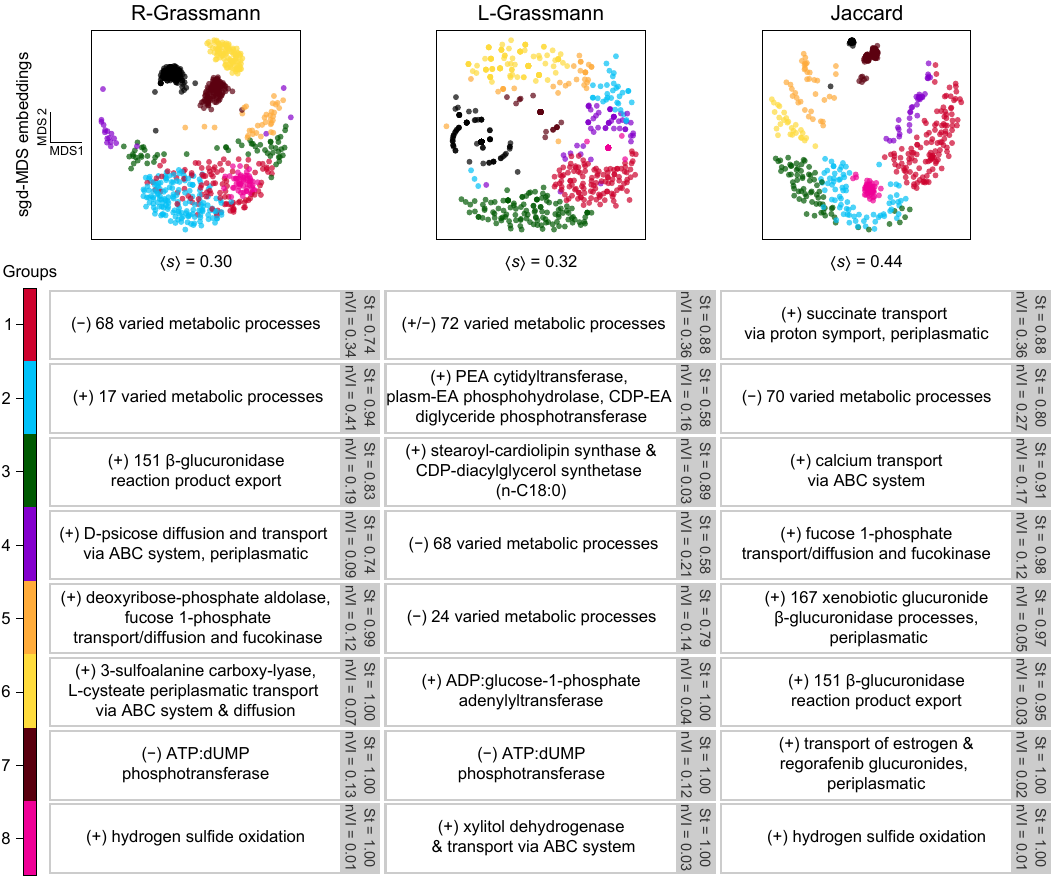}
    \caption{\textbf{Clustering reveals candidate functional groupings enriched or depleted in specific metabolic processes.} 
    Multidimensional scaling embeddings of the metabolic distance matrices are shown colored by clusters obtained by clustering of the matrices with the partitioning around medoids algorithm to produce 8 disjoint clusters. Low-dimensional embeddings are used only for visualization and do not enter the clustering procedure. \textit{Escherichia coli} KOs are shown in black. 
We use normalized variation of information (nVI) to make exploratory annotations of clusters to metabolic processes: zero nVI corresponds to perfect matching, whereas unity nVI corresponds to maximally distinct matching. Cluster out-of-bag stability, or the proportion of bootstrap out-of-bag assignments corresponding to the original cluster, is reported for each cluster. Mean silhouette scores $\langle s \rangle$ are computed for each distance using the eight disjoint clusters as cluster assignments. Abbreviations: ABC = ATP-binding cassette transporters, ATP = adenosine triphosphate, CDP = cytosine diphosphate, CDP-EA = CDP-ethanolamine, PEA = phosphoethanolamine, PlasmEA = plasmenylethanolamine, dUMP = deoxyuridine monophosphate.}
    \label{fig:metabolic-clusterings}
\end{figure*} 

\begin{figure*}[bt!]
\centering
\includegraphics[width=17.8cm]{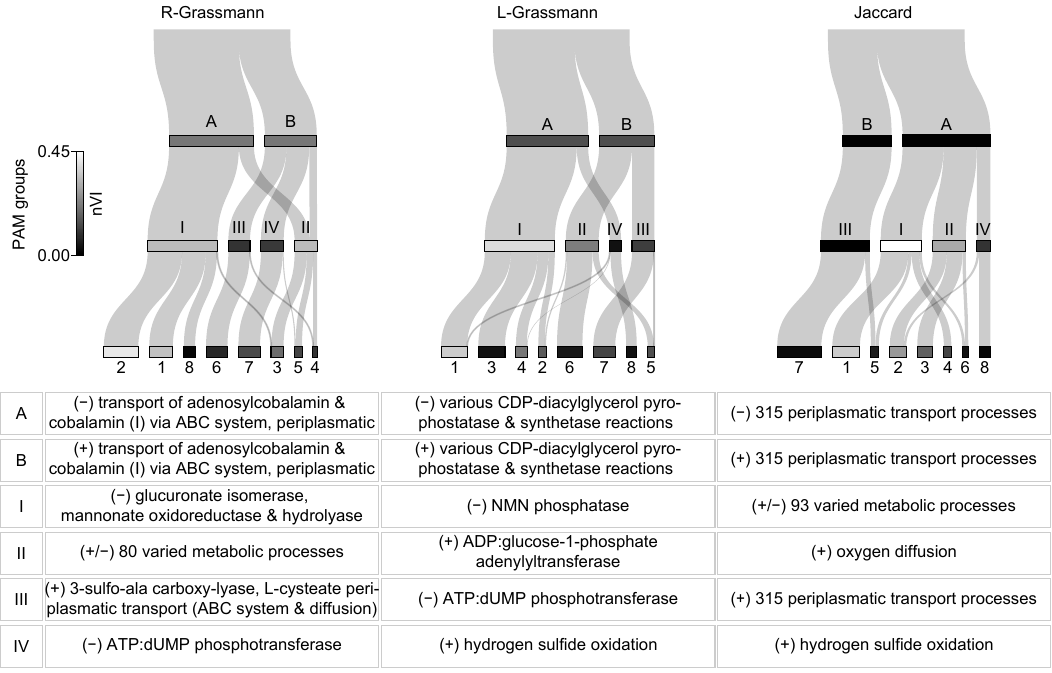}
\caption{\textbf{Effects of granularity on learning candidate functional metabolic groups.} In the spirit of phylogenetic taxonomy, we cluster metabolic distances into 2, 4, and 8 groups [Fig.~\ref{fig:metabolic-clusterings}] using partitioning around medoids. River diagrams show the persistence and splitting of networks in each metabolic group. Correspondingly, the metabolic processes attributed to each group either persist or change demonstrating that, much like classifications in phylogenetic taxonomies, an \textit{a priori} choice of granularity can affect learned functionally informed groupings. Box color and line thickness corresponds to normalized variation of information and number of networks, respectively. Abbreviations: ABC = ATP-binding cassette transporters, ala = alanine, ATP = adenosine triphosphate, CDP = cytosine diphosphate, NMN = nicotinamide mononucleotide, dUMP = deoxyuridine monophosphate.
}
\label{fig:sankey}
\end{figure*}

\section{Embedding and clustering highlights the functional distinguishing power of metabolic distances}

Metabolic processes are the smallest functional subunits of chemical reaction networks. As such, we seek to identify functional metabolic modules that lead to proximal nullspaces and opt for a cluster-based analysis, forgoing associating axes of low-dimensional embeddings to any particular aspect of metabolism. In this manner, we treat cluster annotations as exploratory descriptors rather than formal causal explanations, with the interpretation of metabolic-informed hypotheses that can be developed in future studies. We identify clusters from metabolic distances by  partitioning around medoids (PAM) to the desired level of granularity [Fig.~\ref{fig:metabolic-clusterings}], using the \texttt{R cluster} library. Here, we focus on eight clusters and exclude the \textit{E. coli} KO networks. It should be noted that clusters are not unique and can be interpreted at any resolution. In principle, any desired level of granularity can be considered using this approach. In the spirit of phylogenetic taxonomy~\cite{FelsensteinInferringPhylogenies, Delsuc2005}, we briefly touch upon the coarser cases of two and four clusters in Fig.~\ref{fig:sankey}. 

To assess cluster stability was used an out-of-bag (OOB) resampling procedure. For each of the $B = 200$ resampling iterations, $n$ networks were sampled with replacement from the mini-AGORA2 dataset, where $n$ is the total number of unique networks. The resulting bootstrap distance matrices were clustered using PAM with the same number of clusters as in the original analysis. The medoid indices returned by PAM, which are indexed relative to the bootstrap sample, were mapped back to the corresponding indices in the original dataset. Observations not represented in a given bootstrap sample were designated as OOB observations. For each OOB observation, cluster membership was determined using (i) the nearest medoid from the bootstrap PAM solution and (ii) the nearest medoid from the original PAM solution, based on the same pairwise distance matrix. Stability was evaluated by identifying whether the bootstrap and original nearest medoids belonged to the same cluster in the original PAM partition. An OOB observation was therefore considered stably assigned when the bootstrap and original medoid assignments corresponded to the same original cluster.

For each bootstrap iteration $b$, we recorded $O_b$ which denotes the set of OOB observations and whether each observation $i \in O_b$ was assigned to the same cluster by the bootstrap and original PAM medoids, $\mathbbm{1}[C_i^{\text{boot}}=C_i^{\text{orig}}]$, where $C_i^{\text{boot}}$ is the original-cluster identity of the bootstrap medoid nearest to observation $i$ and $C_i^{\mathrm{orig}}$ is the original-cluster identity of the original PAM medoid nearest to $i$. Cluster OOB stability was then calculated as the proportion of all OOB assignments that were assigned to the same original cluster $c$, 
\begin{equation*}
\operatorname{St}({c}) = \frac{  \sum_{b=1}^B \sum_{k \in O_b }  \mathbbm{1}[ C_k^\text{orig} = c] \mathbbm{1}[C_k^{\text{boot}}=C_k^{\text{orig}}]    }{\sum_{b=1}^B \sum_{k \in O_b }  \mathbbm{1}[ C_k^\text{orig} = c] }.
\end{equation*} 
Thus, cluster OOB stability measures the reproducibility of cluster-specific membership when observations are excluded from the clustering procedure and subsequently assigned using the medoids obtained from the resampled data. 

We considered clusters to be stable if OOB stability was at least 0.80. This threshold corresponds to at least 80\% agreement between cluster assignments obtained from resampled data and those obtained from the original PAM solution when observations were excluded from the clustering procedure. Unsurprisingly, we find that cluster stability varies across metabolic metric and cluster. Thus, we primarily focus on stable clusters, with exceptions for those clusters that may provide additional biological insights.

We identify internal metabolic processes whose presence (or absence) across networks closely matches binary inclusion in a given cluster using variation of information (VI)~\cite{Meilua2003} as a means of quantifying partition similarity. To assign annotations to clusters we perform a cluster comparison analysis for each cluster by masking assignments to inclusion within a cluster of interest and identifying which process(es) are enriched or depleted in the cluster. Since certain processes are always jointly present in the networks, we group processes if they co-occur in the same manner across all our networks. All clusters discussed below are obtained directly from the distance matrices; embeddings are used only for visualization and do not enter the clustering procedure [Fig.~\ref{fig:metabolic-clusterings}. We identify those processes that minimize the variation of information~\cite{Meilua2003} between the inclusion-masked assignments and process-presence. 

For a set $S$, we consider binary partitions $X = \{X_1, X_2\}$ and $Y = \{Y_1,Y_2\}$, where the elements of a partition are disjoint subsets of $S$ whose unions is $S$. We may take $X$ to be a partition based on group inclusion and $Y$ to be a partition based on the presence of co-occurring metabolic processes. VI is defined as~\cite{Meilua2003}
\begin{equation*}
    \text{VI}(X,Y) = H(X) + H(Y) - 2 I(X,Y)
\end{equation*}
where $H$ is the entropy associated with clustering, 
\begin{equation*}
    H(X) = - \sum_{k} P(k) \log P(k), \ P(k) = \frac{\#|X_k|}{\#|S|},  
\end{equation*}
and $I$ is the mutual information between clusterings, 
\begin{gather*}
    I(X,Y) = \sum_{k,k'}P(k,k') \log \frac{P(k,k')}{P(k)P(k')} \\ P(k,k') = \frac{\#|X_k \cap Y_{k'}|}{\#|S|}.
\end{gather*}
For interpretability, we normalize VI by the maximum achievable value for binary partitions: $2 \log 2$. We perform this cluster comparison analysis using the \textsc{Julia Clustering.jl} package~\cite{Bezanson2017}.

In this manner, we identify metabolic processes whose presence/absence patterns align with cluster membership. Although single or co-occurring metabolic processes may be insufficient to fully explain differences in nullspaces, different clusters are nonetheless enriched or depleted in distinct processes [Fig.~\ref{fig:metabolic-clusterings}]. We label the groups R$n$, L$n$, and J$n$ for the R-Grassmann, L-Grassmann, and Jaccard distances, respectively, where $n$ is the group number. To understand the applicability of each metabolic distance in the human gut context, we focus on the clusters with VI below $\sim$10\% of the theoretical upper bound for binary partitions~\cite{Meilua2003}. We identify metabolic clusters for each stoichiometric nullspace [Fig.~\ref{fig:metabolic-clusterings}, left \& center].

Our metabolic cluster analysis identifies annotations shared across distance metrics:  R3 = J6, R5 $\sim$ J4, R7 = L7, and R8 = J8. The R5 and J4 groups are enriched in fucose 1-phosphate transport/diffusion and fucokinase.  Fucokinase is involved in bacterial fucose catabolism which is relevant in the gut as fucose is abundant in host mucin and fucosylated glycans. Fucokinase therefore allows bacteria to use host-derived glycans as carbon and energy sources. The R5 group is additionally enriched in deoxyribose-phosphate aldolase. This reaction belongs to the penotose phosphate pathway and converts 2-deoxyribose-5-phosphate to glyceraldehyde 3-phosphate and acetaldehyde which can then enter central carbon metabolism. R7 and L7 correspond to bacteria lacking ATP:dUMP phosphotransferase, a reaction in nucleotide metabolism that interconverts mono-, di-, and tri-phosphates. Di- and tri-phosphates are significant for their roles in biosynthesis and energy conversion~\cite{Biochemistry}. R8 and J8 are enriched in organisms with the capacity for hydrogen sulfide (H\textsubscript{2}S) oxidation which is notable since H\textsubscript{2}S is known to be redox-active in the human gut~\cite{Barton2017}.

Now let us consider the unique metabolic annotations of each metric. R6 is marked by the presence of a module of L-cysteate utilization which includes a decomposition reaction to taurine and CO\textsubscript{2}. Group L6 is enriched in ADP:glucose-1-phosphate adenylyltransferase, part of the glycogen biosynthesis pathway known to be present in \textit{E. coli}~\cite{EstebanTorres2023}. The L3 group is enriched in stearoyl-cardiolipin synthase and CDP-diacylglycerol synthetase which correspond to a glycerophospholipid metabolism module. In particular, CDP-diacylglycerol synthetase is a core bacterial membrane-biosynthesis reaction that generates CDP-diacylglycerol \cite{Blunsom2020}. In bacteria, CDP-diacylglycerols are precursors for the biosynthesis of all major phospholipids, including phosphatidylglycerol and cardiolipin, that comprise organelle membranes \cite{Biochemistry, Blunsom2020}. Stearoyl-cardiolipin synthase converts two phosphatidylglycerol molecules into cardiolipin and glycerol, contributing to bacterial membrane phospholipid composition \cite{Hirschberg1972}. Cardiolipins are important membrane lipids involved in membrane architecture and the function of membrane-associated proteins \cite{Gottier2017}. Thus, these processes primarily reflect bacterial membrane maintenance rather than a host-specific function. Group L8 is enriched for processes associated with a xylitol utilization module, consisting of ATP-dependent transport by an ABC transporter and oxidation of xylitol to xylulose by xylitol dehydrogenase \cite{Jefferies1983}. In the gut, this capacity may be relevant since xylitol is a dietary sugar alcohol that is incompletely absorbed and can reach the colon \cite{Flint2012}, where microbes capable of xylitol utilization may ferment it and participate in niche-specific carbohydrate metabolism \cite{Koropatkin2012}. 

The R-Grassmann and Jaccard metrics produce clusters characterized by glucuronide-related processes [Fig.~\ref{fig:metabolic-clusterings}]: R4, R6, J1, J5, and J7. In particular, J5 and J7 corresponds to the presence of periplasmatic transport processes for metabolites conjugated to glucuronate, including the three most abundant estrogens---estradiol, estriol, and estrone---and the drug regorafenib~\cite{Thomas2013}. Glucuronide moieties are added to metabolic substrates to increase hydrophilicity and thereby facilitate elimination from the human body. Glucuronidation is therefore critical for the removal of unwanted endogenous molecules, drugs, and xenobiotics~\cite{Blanco1985,Yang2017,Gao2022}. 
Additionally, group R4 is enriched in D-psicose, more commonly known as D-allulose, perplasmatic diffusion and transport via ATP-binding cassette transporters that couple ATP hydrolysis to the influx and efflux of substrates~\cite{Davidson2004}. Group J1 is enriched in the periplasmatic transport of the metabolite succinate via a proton symport. The presence of periplasmatic annotations is likely an indication of clusters rich in Gram-negative organisms \cite{Hobot1984, Matias2006}. Both Grassmann distances also produce clusters over-represented in \textit{E. coli} strains: R6-7 and L6-7. The redundancy reported here indicates that clusters and annotations may persist across metrics. In the case of \textit{E. coli} strain-rich clusters, this is due to an overabundance of similar networks. Otherwise, this redundancy may be due to similar over-representations or a biological signal invariant to the choice of metric. Nonetheless, as this is an inherent drawback of cluster-based annotations rather than the choice of metric, further validation is required. 

In our cluster analysis, we observe that the Jaccard distance largely does not annotate the same metabolic processes as the Grassmann distances at any level of cluster granularity [Figs.~\ref{fig:metabolic-clusterings} \& \ref{fig:sankey}]. Although clustering is heavily dependent on data, all metabolic distances and clusters considered here are computed on the same dataset. The absence of cluster annotation agreement between the Jaccard distance and Grassmann distances suggests that, for the microbial metabolic networks considered here, binary comparisons are insufficient to identify differences in resource utilization and conserved metabolic pools. More generally, Grassmann and Jaccard distances capture different aspects of metabolic organization: Jaccard distances emphasize metabolic process-overlap, whereas Grassmann distances incorporate stoichiometric geometry through the unconstrained steady-state flux and conservation-law subspaces.

While the theoretical bound is not saturated for any cluster, the observed VI values suggest that the biological basis of these groupings may depend on combinations of processes rather than any single metabolic process alone; systematically resolving such combinations will likely prove to be combinatorially prohibitive for large sets. Overall, these results support that nullspace geometry captures biologically meaningful variation beyond binary content. By incorporating stoichiometric structure and conserved metabolic degrees of freedom, the Grassmann distances resolve differences in metabolic organization that are largely invisible to Jaccard overlap alone (Appendix~\ref{sec:chemoton}).

\section{Phenotype prediction highlights the predictive power of metabolic distances}

To assess the functional distinguishing power of metabolic and phylogenetic distances, we examined the prediction of external phenotype annotations obtained from the JGI IMG/M~\cite{Chen2023} in a supervised LOOCV framework. We considered seven binary phenotypes (Supplemental Table 1): oxygen requirement (anaerobe versus facultative), Gram staining, motility, pathogenicity, sporulation ability, location (oral versus intestine), and cell shape (rod- versus coccus-shaped). The IMG/M annotations are sparse and imbalanced, as well as over-represented for well-studied organisms. Additionally, these binary phenotypes may be encoded in the mini-AGORA2 networks by the curation of specific metabolic 

\begin{figure*}[bt!]
\centering
\includegraphics[width=17.8cm]{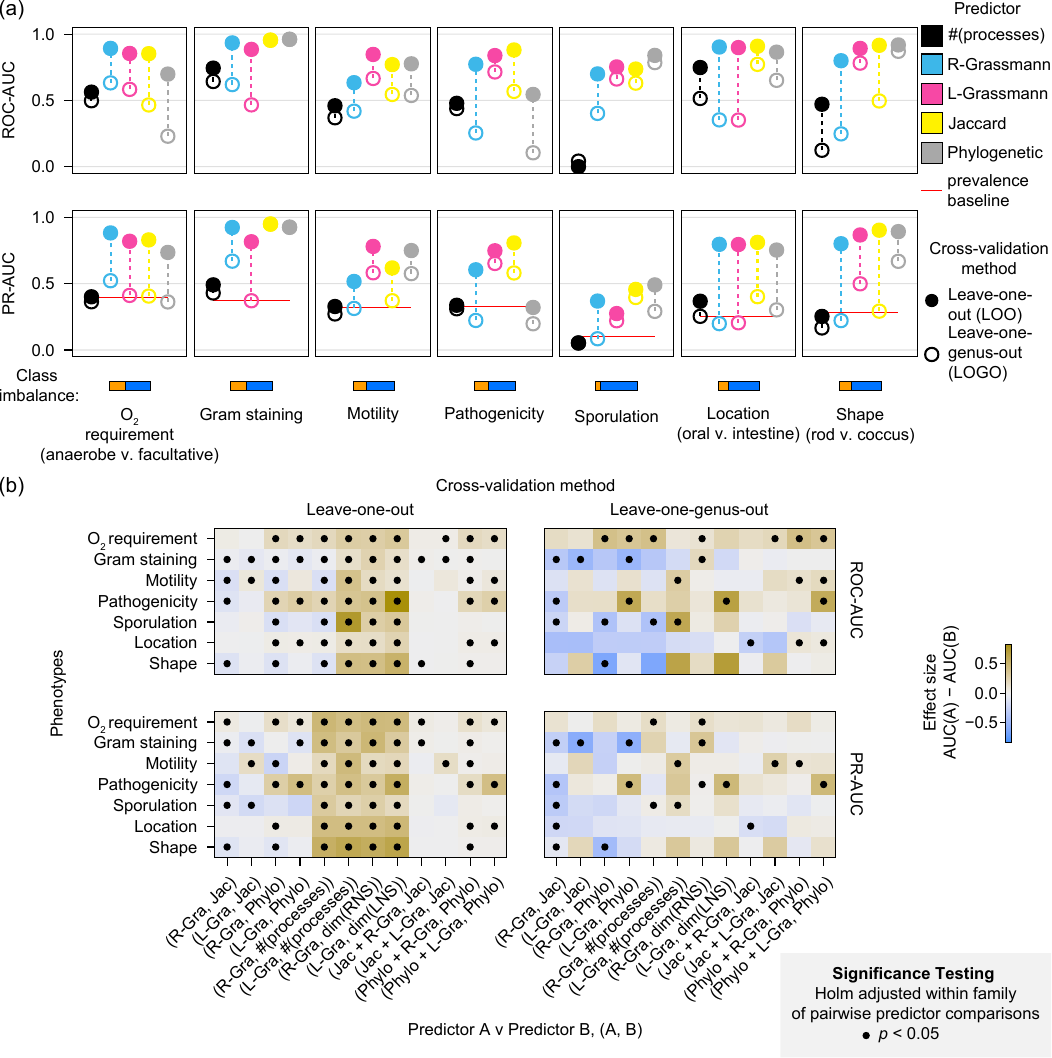}
    \caption{\textbf{Metabolic distances differentially distinguish microbiological phenotypes.}
    (a) Leave-one-out  and leave-one-genus-out cross-validated (top) ROC-AUC and (bottom) PR-ROC for prediction of seven binary bacterial phenotypes using signed distance-to-class-prototype scores. Scores are computed from L-Grassmann, R-Grassmann, and Jaccard distances of metabolic networks, as well as from phylogenetic distances and the number of metabolic processes. AUC performance and 95\% confidence intervals are reported in Supplemental Tables 2 and 3. Additionally, Brier scores (mean-squared error for predicted probabilities) and log loss values are reported in these tables. The smallest class is taken to be the positive class with the class imbalance as shown. ROC-AUC is insensitive to class imbalance but may be over-optimistic in a class-imbalanced setting. As such the PR-AUC is also reported (Middle) with the PR baseline is shown in red and corresponds to the fraction of positive class networks. 
    (b) The effect sizes in the AUC values resulting from changing predictors within the LOO or LOGO framework are shown. Statistical significance ($p < 0.05$) was determined from a paired bootstrap test with finite-sample correction. $p$-values are Holm adjusted within each family of pairwise metric comparisons. In the LOO setting, Grassmann predictors for specific external phenotypes (oxygen requirement and motility), and all metabolic distances outperform the simple size-based baseline, suggesting that Grassmann geometry captures biologically meaningful structure in bacterial metabolic networks. }
    \label{fig:phenotype-prediction}
\end{figure*}
\vfill
\clearpage
\newpage
\noindent processes during model reconstruction and refinement. As such, biased data curation is an expected limitation of this assessment. However, the differing ability of each metric to predict these binary phenotypes is also related to the properties encoded by the metrics. As the unique networks show the same binary phenotypes as their duplicates, we used the unique mini-AGORA2 networks to avoid data leakage. 

We evaluated predictive performance using two cross-validation protocols. In leave-one-out cross-validation (LOOCV), each network was held out once whereas in leave-one-genus-out cross-validation (LOGOCV), all networks belonging to one genus were held out together. In either protocol, class prototypes, score-standardization, and Firth regression were computed exclusively from the outer training set. Folds in which the training set did not contain both phenotype classes were not evaluated by regression and set to the training-fold prevalence baseline. For each phenotype-prediction task, we restricted the analysis to organisms with one of the two relevant phenotype labels and fixed the less prevalent class as the positive class for all analyses and cross-validation folds. Organisms with identical metabolic networks and phenotype labels were collapsed to a single representative network to prevent data leakage.  Otherwise, the protocols proceed as in Section~\ref{sec:GENETIC-VARIATION} [Fig.~\ref{fig:ecoli-ko-viability}] with signed score 
\begin{equation*}
   \operatorname{score}(\mathcal{N}_i) = \dist(\mathcal{N}_i, \mathcal{N}_\text{0-prototype}) -  \dist(\mathcal{N}_i, \mathcal{N}_\text{1-prototype}). 
\end{equation*}
To define a class prototype for each $\sigma$ and within each fold, we compute the Fr\'{e}chet medoid which is defined as
\begin{equation*}
    \mathcal{N}_{\sigma\text{-prototype}} = \underset{\mathcal{N}_k \in \mathcal{S}_\sigma }{\argmin}  \sum_{\mathcal{N}_j \in  \mathcal{S}_\sigma} \dist(\mathcal{N}_k, \mathcal{N}_j) . 
\end{equation*}
where $\mathcal{S}_\sigma$ is the set of the networks in the $\sigma$ class. The Fr\'{e}chet medoid class prototype is simply the network that minimizes that total distance to other organisms in its class. For the phylogenetic distance, each set of organisms with identical metabolic networks was represented by its Fr\'{e}chet medoid in phylogenetic-distance space. As simple baselines, we also considered the signed scores obtained by computing pairwise absolute differences in the number of metabolic processes or nullspace dimension. When a set contained only two equally valid medoids, the first organism in lexicographic order was selected.

For each predictor and phenotype we estimated out-of-fold probabilities with Firth logistic regression models fitted using only the corresponding outer training data. Predictor scores were standardized using training-fold statistics before model fitting, and the same transformations were applied to the held-out network or genus. Predictive performance was then calculated from the pooled out-of-fold probabilities, such that every eligible network contributed exactly one probability under each predictor and cross-validation protocol. We evaluated both discrimination, using ROC-AUC and PR-AUC, and probabilistic accuracy, using log loss and Brier score. All values are reported in Supplemental Tables 2 \& 3. 

We first evaluated each predictor separately using models of the form of Eq.~(\ref{eq:SIMPLE-LOGISTIC-PREDICTOR}) where the training-fold-standardized scores are derived from the R-Grassmann, L-Grassmann, Jaccard, phylogenetic, number of metabolic processes, or nullity predictor. The single-predictor results indicate that phenotype prediction depends on both the network descriptor and the validation protocol [Fig.~\ref{fig:phenotype-prediction}(a)]. Under LOOCV, several metabolic and phylogenetic predictors produced strong discrimination. Performance generally decreased under LOGOCV, consistent with the greater difficulty of predicting phenotypes for organisms from genera excluded from model training. This decrease was not uniform across phenotypes or predictors, however, indicating that some associations generalized beyond genus-level similarity more effectively than others. Consistent with the LOOCV analysis, the clearest phenotype-specific Grassmann advantages were observed for oxygen requirement and motility. R-Grassmann was particularly informative for distinguishing anaerobic from facultative organisms, whereas L-Grassmann was particularly informative for distinguishing motile from nonmotile organisms. These associations persisted to varying degrees under LOGOCV, supporting the conclusion that the corresponding nullspace geometries contain phenotype-related information that is not restricted to comparisons among organisms from genera represented during training. By contrast, Jaccard and phylogenetic predictors remained competitive or superior for several other phenotypes. Thus, the results do not support uniform superiority of Grassmann distances; rather, their predictive value is phenotype and nullspace dependent.

To distinguish direct predictive superiority from complementary information, we performed two types of model comparison [Fig.~\ref{fig:phenotype-prediction}(b)]. First, we compared the out-of-fold performance of separately fitted single-predictor models. For a pair denoted $(A,B)$, the displayed effect size is
\begin{equation*}
\Delta \text{AUC}(A,B)=\text{AUC}(A)-\text{AUC}(B),
\end{equation*} 
where AUC is either ROC-AUC or PR-AUC. Positive values therefore indicate better discrimination by predictor $A$, whereas negative values favor predictor $B$. These comparisons included R-Grassmann and L-Grassmann against Jaccard, phylogenetic distance, number of metabolic processes, and the corresponding nullspace dimension. Comparisons with the number of metabolic processes and nullity assess whether Grassmann-based prediction can be explained by simple differences in network size or nullspace dimension.

Second, we fitted nested Firth logistic models to test whether a Grassmann predictor added information to Jaccard or phylogenetic predictors. The comparison {(A + B, A)} contrasts the augmented model
\begin{equation*}
\operatorname{logit}P(Y=1) = \beta_0+\beta_Az_A+\beta_Bz_B
\end{equation*}
with the reduced model
\begin{equation*}
\operatorname{logit}P(Y=1) = \beta_0+\beta_Az_A. 
\end{equation*}
Analogous comparisons were performed by adding L-Grassmann to Jaccard and by adding either Grassmann predictor to phylogenetic distance. All score standardization and coefficient estimation were performed independently within each outer training fold. Consequently, positive nested-model effect sizes indicate that adding the Grassmann predictor improved out-of-fold discrimination relative to the reduced comparator model. The nested comparisons provide a more stringent test of complementarity than head-to-head predictor comparisons. A Grassmann predictor can be less accurate than Jaccard or phylogenetic distance when used alone. However, a prediction task may improve when included jointly if a distinct component of phenotype-associated variation is captured by the Grassmann predictor. Conversely, strong performance by a Grassmann predictor alone does not establish incremental information if its addition fails to improve the corresponding reduced model.

Bootstrap $p$-values were calculated as in Section~\ref{sec:GENETIC-VARIATION} [Fig.~\ref{fig:ecoli-ko-viability}] with boostrapping of networks for LOOCV or genera for LOGOCV. $p$-values were then adjusted using Holm’s method within each prespecified family of predictor comparisons, i.e. R-Grassmann versus Jaccard, L-Grassmann versus Jaccard, R-Grassmann versus phylogenetic, and so forth. Black markers in Fig.~\ref{fig:phenotype-prediction}(b) indicate comparisons with Holm-adjusted $p<0.05$. Because statistical significance depends on both effect magnitude and uncertainty, the full effect estimates, bootstrap confidence intervals, unadjusted $p$-values, and Holm-adjusted $p$-values are reported in Supplemental Tables 2 \& 3. 

Direct model comparisons and $p$-values show that selected Grassmann advantages are statistically significant and persist relative to Jaccard distance, phylogenetic distance, number of metabolic processes, and the corresponding nullspace dimension [Fig.~\ref{fig:phenotype-prediction}(b)]. Thus, the predictive signal is not fully explained by network size or nullity. Nested comparisons further show that, for specific phenotypes, adding a Grassmann predictor improves out-of-fold discrimination relative to a model containing Jaccard or phylogenetic distance alone. These results support the interpretation that stoichiometric nullspace geometry provides complementary phenotype-related information rather than a uniformly superior replacement for Jaccard or phylogenetic descriptors.

ROC-AUC and PR-AUC evaluate discrimination but do not determine whether the predicted probabilities are accurate. We therefore evaluated probabilistic performance using out-of-fold log loss and Brier score, with smaller values indicating better predictions. The same held-out probabilities used for the discrimination analysis were used for these proper scoring rules. As in the case of the \textit{E. coli} KO viability prediction [Fig.~\ref{fig:ecoli-ko-viability}], differences between discrimination and probabilistic performance indicate that a predictor may rank organisms correctly without producing equally accurate probability estimates. 

We also repeated the phenotype analysis under alternative representations of the stoichiometric networks. For R-Grassmann, we examined $\ell^1$ and $\ell^2$ column-wise normalized stoichiometric matrices (Supplemental Table 4); for L-Grassmann, we examined the left nullspaces of the corresponding closed networks (Supplemental Table 5). Each sensitivity analysis repeated the full cross-validation procedure, including fold-specific score standardization, Firth model estimation, and prediction of held-out probabilities. The normalization results assess dependence of R-Grassmann prediction on the coordinate convention, whereas the closed-network results assess whether phenotype prediction is retained when the left nullspace describes latent internal conservation structure after exchange processes are removed. The ROC-AUC and PR-AUC R-Grassmann advantage over Jaccard persists after $\ell^1$- and $\ell^2$ column-wise normalization, with significance vanishing only after Holm adjustment for ROC-AUC discrimination (Supplemental Tables 2 \& 3). In the same context, the R-Grassmann advantage against the phylogenetic distance is preserved. Clearly, the choice of normalization impacts the statistical significance of the advantage against Jaccard, but $\ell^1$ and $\ell^2$ are not sufficient to ablate the effect size. Moreover, the ROC-AUC and PR-AUC advantage of the L-Grassmann distance over Jaccard and phylogenetic distances in predicting organism motility is lost after network closure, suggesting that the latent conservation structure of metabolic networks is insufficient to accurately discriminate motile from nonmotile organisms. The discriminatory information is therefore associated with the conservation relations that survive under the open-network boundary, rather than with latent internal conservation structure alone. This association may reflect biologically meaningful differences in environmental exchange, differences in the reconstruction and annotation of exchange reactions, or both.

Taken together, these analyses show that stoichiometric nullspace geometry contains out-of-sample predictive information for selected organism-level phenotypes. Direct comparisons establish where Grassmann-based models outperform competing single-predictor models, while nested comparisons assess whether Grassmann predictors add information beyond network overlap or phylogenetic relatedness. The LOGOCV results further determine which associations generalize to organisms from genera absent during model training. The absence of uniform improvements across phenotypes underscores that the predictive relevance of nullspace geometry depends on the biological trait and network representation. Grassmann predictors should therefore be regarded as physically interpretable complements to Jaccard, network-size, nullity, and phylogenetic descriptors rather than as replacements for them.\vfill

\section{Grassmann distances on human tissue and planetary chemical reaction networks} 

\begin{figure*}[bt!]
\centering
\includegraphics[width=17.8cm]{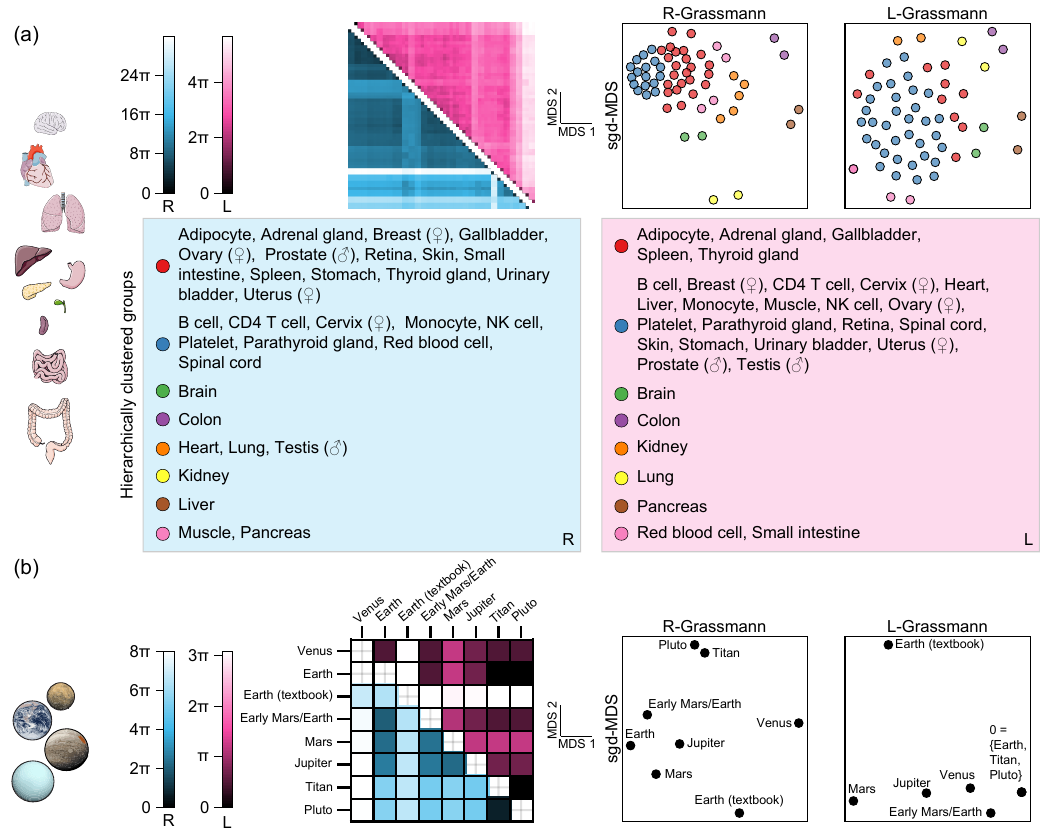}
    \caption{\textbf{Comparisons of human tissues and planetary atmosphere highlights applicability of  metabolic Grassmann distances to other complex chemical reaction networks.} (a) Stochastic gradient descent multidimensional scaling embeddings of metabolic Grassmann distance computed on sex-specific human tissue and organ metabolic networks are shown colored by 8 clusters obtained by hierarchical clustering of the matrices with Ward's linkage and a tree-cut to produce 8 disjoint clusters for visualization purposes. Graphical elements were adapted from Servier Medical Art under a Creative Commons Attribution 4.0 license. 
    (b) Stochastic gradient descent multidimensional scaling embeddings of metabolic Grassmann distance computed on 8 published planetary atmosphere chemical reaction networks serve as a proof of concept based on limited data availability. As additional planetary networks becomes available, these can be incorporated. Distance matrix labels are in order of distance from the Sun. Human tissue and planetary atmosphere chemical reaction networks were obtained from~\cite{Thiele2020} and~\cite{Wong2023}, respectively. Planetary graphical elements were obtained from the National Aeronautics and Space Administration.}
    \label{fig:organs-planets}
\end{figure*}

The nullspace-based framework presented here is general and can be applied to complex chemical reaction networks on various length scales. As illustrative examples, we consider the chemical reaction networks of sex-specific human tissues~\cite{Thiele2020} and planetary atmospheres~\cite{Wong2023}. Despite metabolic differences between tissues of different sexes, we find that the Grassmann distances group networks by tissue type [Fig.~\ref{fig:organs-planets}(a)]. We also observe clustering based on hematopoietic-stem cell origin (B cells, CD4 T cells, monocytes, NK cells, platelets, and red blood cells)~\cite{Weissman2001}. We note that the tissue analysis is intended to illustrate applicability across multicellular metabolic reconstructions rather than to provide a comprehensive classification of human tissue metabolism.

\begin{figure*}[t!]
\centering
\includegraphics[width=17.8cm]{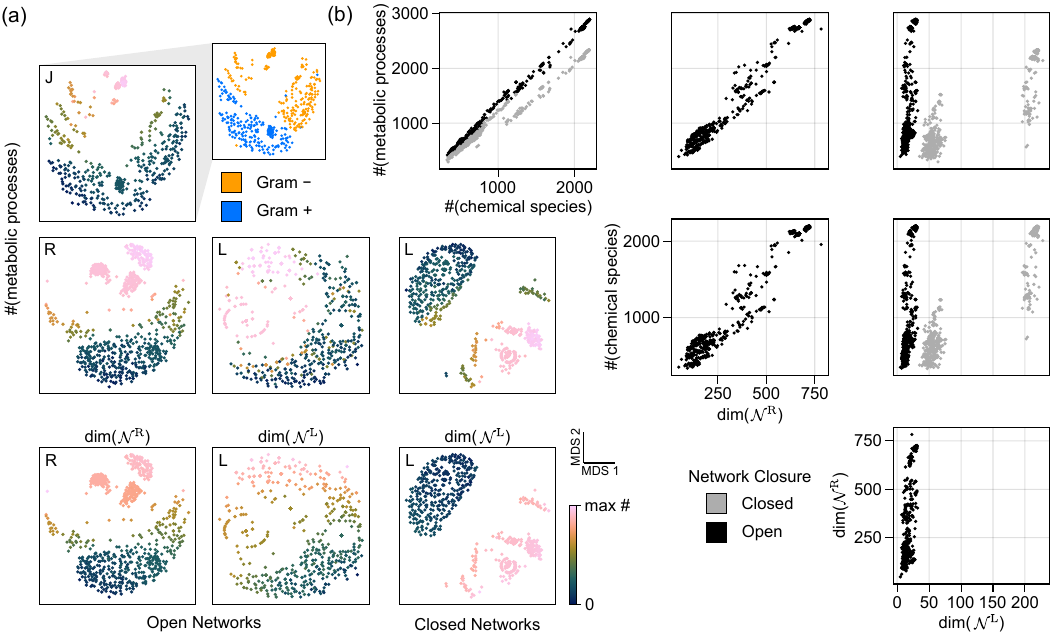}
\caption{\textbf{Dimension of nullspace colors direction in Grassmann embeddings.} (a) Colorings the sgd-MDS embedding of the Grassmann distance matrix by nullspace dimension and network size are consistent with directions corresponding to nullspace size. The color scheme of individual embeddings range from 0 to the maximum number observed for either the number of metabolic processes or nullity. Embeddings of R-Grassmann, L-Grassmann, and Jaccard distances are labeled with an R, L, and J, respectively. Gram-staining information is shown for the Jaccard embedding as an inset highlighting that the largest networks are Gram negative. (b) Bacterial network size is linearly predictive of the nullspace dimension in open networks, in contrast to closed networks.}
\label{fig:size-nullity}
\end{figure*}

To move towards chemical reaction networks on the astrophysical scale, we consider eight atmospheric networks corresponding to six celestial bodies [Fig.~\ref{fig:organs-planets}(b)]. The networks used here are derived from a study on the chemical reaction networks of planetary atmospheres using graph topological measures~\cite{Wong2023}. With the exception of the ``textbook'' Earth model~\cite{Sole2004}, these networks were obtained from published studies simulating the photochemistry of atmospheres using the photochemistry–transport model KINETICS from Caltech–JPL~\cite{Allen1981}. Importantly, we observe that the most distal network corresponds to the ``textbook'' Earth atmospheric model, which suggests network curation effects on Grassmann distances. Because the available atmospheric reaction networks are few and heterogeneous in curation depth, we regard this analysis as a proof of concept rather than a robust classification of planetary atmospheres. However, as more networks become available, these can be readily incorporated.  

\section{Sensitivities and limitations of the Grassmann distance}

The Grassmann distance is definitionally susceptible to primarily recapitulate differences in subspace dimension and network size (Appendix~\ref{sec:dist-components}). We empirically examine the extent of this susceptibility for Grassmann distances between mini-AGORA2 networks. Moreover, as with any data-driven computational analysis, data quality and curation have the potential to affect conclusions. We discuss potential observations of such effects in our analyses. 

\subsection{The role of network size and nullity on the Grassmann distance}

The Grassmann distance is composed of two components: a dimension gap and an angular term. From the rank-nullity theorem, we know that the dimension gap component does not only capture differences in network size, but also differences in stoichiometric degrees of freedom (Appendix~\ref{sec:dist-components}). Moreover, the Grassmann distances computed on the mini-AGORA2 dataset are not dominated by only the dimension gap [Fig.~\ref{fig:size-effects}(a)]. Nonetheless, we expect and find that the Grassmann distances are correlated with differences in nullity and network size [Fig.~\ref{fig:size-effects}(b)-~\ref{fig:size-effects}(c)]. Additionally, on the 8 clusters of Fig.~\ref{fig:metabolic-clusterings}, we find that within-cluster distances have a substantial angular term contribution, whereas between-cluster distances are mostly due to the dimension gap. Even so, there is no clear domination by either Grassmann component at this cluster resolution. Furthermore, the phenotype-prediction performance of the Grassmann distance significantly exceeds a simple network size baseline [Figs.~\ref{fig:ecoli-ko-viability} \& \ref{fig:phenotype-prediction}] in a nullspace and predictor dependent manner. Together, the Grassmann distance goes beyond capturing differences in network size and nullity. 

\subsection{Effects of network curation on metabolic distances}

Any data-driven computational analysis is limited by data availability; chemical reaction networks are no different. In practice, genome-scale reconstructions of metabolism are limited by uncertainties present in the reconstruction pipeline such as incomplete or missing gene annotations~\cite{Bernstein2021}. The authors of AGORA2~\cite{Heinken2023} addressed these concerns using a semiautomated refinement pipeline that curatorially adds ``missing'' metabolic processes using experimental data and removes thermodynamically infeasible processes~\cite{Heinken2021, Heinken2023}. Additionally, most of the planetary atmosphere networks examined here lack ``textbook-level'' detail which may explain the distant textbook Earth network in Fig.~\ref{fig:organs-planets}(b). Rather, these networks were constructed to reproduce known physical parameters and sparse chemical data~\cite{Wong2023}. Consequently, we caution against overinterpretations of distances between chemical reaction networks. Nonetheless, further well-curated atmospheric network datasets will enable more meaningful comparative studies.

For the microbial networks considered here, the processes identified for R4, R6, J1, J5, and J7
occur across the periplasm, the region between the outer and inner cell membrane in bacteria with two membranes (diderms) or the region between the cell membrane and the cell well in bacteria with a single membrane (monoderms)~\cite{Hobot1984, Gupta1998, Matias2006, Sutcliffe2010, Tocheva2016}. In general, monoderms are Gram-positive and diderms are Gram-negative; however, there are notable exceptions~\cite{Sutcliffe2010, Tocheva2016}. Previous work has shown that Gram stain reactivity of organisms produces well-defined clusters with the Jaccard distance~\cite{Bauer2015}. We recapitulate this result with the Jaccard distance which significantly outperforms the Grassmann distances in predicting the Gram stain phenotype [Fig.~\ref{fig:phenotype-prediction}]. We note that the Gram-negative network set in mini-AGORA2 contains the largest mini-AGORA2 networks, and this is consistent with the performance of the baseline predictor based on the number of processes. [Figs.~\ref{fig:phenotype-prediction} \& \ref{fig:size-nullity}(a)]. Is it then surprising that the Jaccard distance, a metric based on set overlap, produces well-defined clusters on the basis of Gram-stain reactivity?  We also note that AGORA2 networks were curated to include drug biotransformation and degradation reactions to enable the modeling of personalized gut microbial drug metabolism~\cite{Heinken2023}. That we appreciably observe this feature when clustering with the Grassmann and Jaccard distances further underscores the need for a nuanced and context-aware interpretation of metabolic distances. 

\section{Beyond optimal metabolic adaption}

Experimental work has shown that deletion of the \textit{pyk} gene in \textit{E. coli} JM101 leads to local redistribution of metabolic fluxes~\cite{Emmerling2002}. Subsequent \textit{in silico} KOs of \textit{E. coli} genome-scale metabolic reconstructions showed local redistributions of steady-state flux vectors obtained by minimization of metabolic adjustment (MOMA). MOMA matches an element of the right nullspace in a genetically perturbed metabolic network to a flux balance analysis (FBA) solution of the wild-type (WT) metabolic network that minimizes the difference in fluxes, $\lVert v_\text{WT} - v_\text{KO} \rVert$, subject to additional flux constraints. MOMA fluxes are consistent with the hypothesis that laboratory KOs need not satisfy optimal metabolic adaptation~\cite{Segre2002}. Here we find a stronger statement: nearly all computationally viable gene KOs of \textit{E. coli} K-12 MG1655 do not produce appreciable differences in the spaces of unconstrained steady-state fluxes and conservation laws compared to the variation seen in \textit{E. coli} strains. Therefore, within the reconstruction and viability criteria used here, many single-gene perturbations leave the nullspace structure of the network approximately unchanged. These results suggest that conservation laws and unconstrained steady-state network fluxes critical for biological function are insulated from genetic perturbations. We note that this may aid in the design of synthetic organisms that recapitulate these salient metabolic features from wild nonsynthetic organisms~\cite{Yu2002,Aydin2022, Moger-Reischer2023, Mardoukhi2024}. Similarly, we should not be surprised if these principles are applicable to the search for Earth-like planetary atmospheres where specific steady-state fluxes and conservation laws may be critical for sustaining life~\cite{Wong2023}. 

\begin{figure*}[bt!]
\centering
\includegraphics[width=17.8cm]{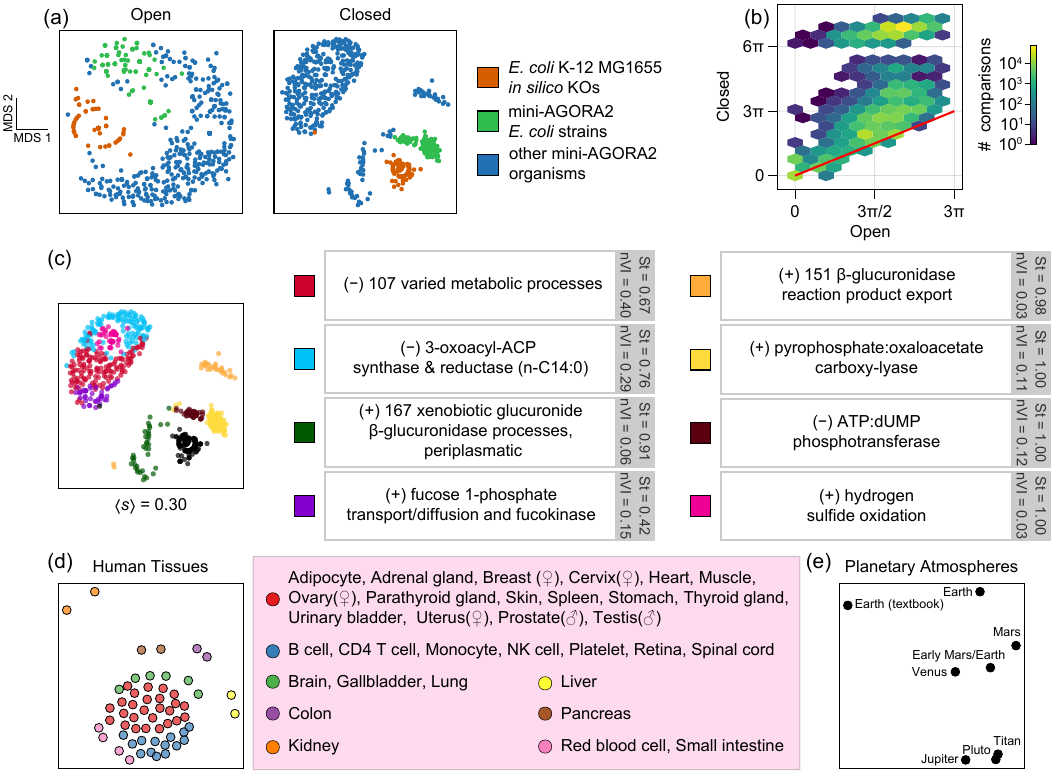}
\caption{\textbf{Closed chemical reaction networks serve as auxiliary descriptors of latent conservation structure.} (a) sgd-MDS embeddings of closed bacterial metabolic networks show improved separation of non-\textit{E. coli} networks from \textit{E. coli} networks. (b) Distances between open networks are generally smaller than their closed counterparts. (c) Our cluster comparison analysis on the L-Grassmann distance of closed bacterial metabolic networks using paritioning around medoids leads to the identification of metabolic processes not identified in open networks. Conservation laws in human tissues (d) and planetary atmospheres (e) are also affected by network closure as illustrated by sgd-MDS embeddings and hierarchical clustering with Ward's linkage for visualization purposes.
     Abbreviations: ACP = acyl carrier protein, ATP = adenosine triphosphate, dUMP = deoxyuridine monophosphate.}
    \label{fig:open-vs-closed}
\end{figure*}

\section{Open vs closed chemical reaction networks} 

Until now, our primary analyses have used open chemical reaction networks for both nullspaces, so that the right and left nullspace distances are defined on the same stoichiometric matrix. Here, we examine how the results change when the left nullspace is instead computed after network closure. We find that L-Grassmann distances on closed networks generally exceed their open-network counterparts [Fig.~\ref{fig:open-vs-closed}(b)]; $< 1\%$ of open-network distances are strictly larger. We notably observe this in the comparison of the planetary atmosphere networks, where three open networks (Earth, Titan, Pluto) have a trivial left nullspace and consequently map to the same point in a MDS embedding [Fig.~\ref{fig:organs-planets}(b)]. This is not a limitation of the metric, but rather a property of the network and boundary; networks with trivial stoichiometric left nullspaces are able to explore the full concentration space of their chemical species. Consistent with an increase in the number of conservation laws, this degeneracy is remedied by the removal of exchange processes [Fig.~\ref{fig:open-vs-closed}(e)]. We also observe changes to the L-Grassmann distances of the microbial and human tissue networks [Figs.~\ref{fig:open-vs-closed}(a)-\ref{fig:open-vs-closed}(d)] and the subsequent analysis of metabolic groups (Appendix~\ref{sec:open-network-groups}).

We emphasize that the closed-network left nullspace should not be interpreted as the conservation law structure of the physically open system itself. Rather, it is an auxiliary descriptor of the corresponding internally closed network. Biological systems are phenomenologically open; energetic and material demands for maintenance and growth are satisfied by exchange with the environment~\cite{Polettini2014}. However, systems with the potential to take up materials from the environment may not do so at every point in time. Therefore, the open-network left nullspace is the more direct physical object, while the closed-network left nullspace can be useful for quantifying latent conservation structure independent of exchange. The extent to which a partially open-network representation may provide a more faithful intermediate description is a question for future work. 

\section{Conclusions}

Recent experimental and computational developments provide us with the opportunity to develop functional classifications of chemical reaction networks grounded in physical principles. Here, we introduce a framework for comparing chemical reaction networks using differences between nullspaces of their stoichiometric matrices. In the human gut microbiome, the framework identifies geometric organization in stoichiometric nullspaces that is distinct from phylogeny and predictive of several external phenotypes. While the Grassmann distance over the stoichiometric nullspaces is not uniformly superior to the Jaccard or phylogenetic distances in phenotype prediction, it provides complementary geometric descriptors whose predictive value depends on the phenotype, nullspace, and taxonomic generalization setting.

The generality of this framework, from chemical reaction networks in bacteria to human tissues to planetary atmospheres, can lead to the development of a universal atlas of chemical reaction networks in which systems across length scales must reside. In this broader setting, Grassmann distances should be viewed not as replacements for phylogenetic metrics, or other metrics on underlying processes, and network-overlap metrics, but as complementary, physically interpretable descriptors of stoichiometric organization that can reveal functional variation not captured by those comparisons alone. Moreover, by recasting complex nonlinear dynamical systems as effective transport networks in the form of Eq.~(\ref{eq:MASS-ACTION}), the above  methodology becomes broadly applicable to other dynamical phenomena.

\section*{Data Availability} 
All study data and relevant codes are available on a public Github repository at \url{https://github.com/jrysyrj}.
\section*{AI Use Statement}
The derivation of the rescaling bound shown in Appendix A3 was developed with the assistance of OpenAI's GPT-6 Luna as a querying and generation tool; all steps in the derivation were verified to be correct by the authors. 

\section*{Acknowledgments} 
We are grateful to Gene-Wei Li and Leonid Mirny for helpful discussions. We thank Stas Shvartsman for introducing us to the chemoton. We also acknowledge the MIT SuperCloud, Lincoln Laboratory Supercomputing Center, and  MIT Office of Research Computing and Data for providing high performance computing resources that have contributed to the research results reported within this paper. This work was supported by a MathWorks Science Fellowship (J.R.), the National Science Foundation Graduate Research Fellowship Program under Grant No. 2141064 (J.R.), the National Science Foundation DMR/MPS-2214021 (J.D.), the MathWorks Professorship Fund (J.D.), Alfred P. Sloan Foundation Grant G-2021-16758 (J.D.), and through Schmidt Sciences LLC (Polymath award to J.D.). Any opinions, findings, and conclusions or recommendations expressed in this material are those of the author(s) and do not necessarily reflect the views of the National Science Foundation.

\appendix
\section{STOICHIOMETRIC NULLSPACES \label{sec:stoich-nullspaces-transport-networks}}

Here we clarify the physical interpretation of the stoichiometric right and left nullspaces and the role of network closure. For a chemical reaction network with dynamics $\dd c/\dd t=Sv$, the right nullspace contains unconstrained steady-state flux combinations satisfying $Sv=0$, whereas the left nullspace contains linear combinations of species satisfying $w^\top c=\mathrm{constant}$. Because exchange processes alter the system boundary, they can break conservation laws and thereby change the left nullspace. The examples below make these statements explicit and motivate the distinction between the open-network comparisons used in the main text and the auxiliary left nullspace analysis on closed networks.

\subsection{Developing a physical interpretation}
 
\begin{figure*}[bt!]
\centering
\includegraphics[width=17.8cm]{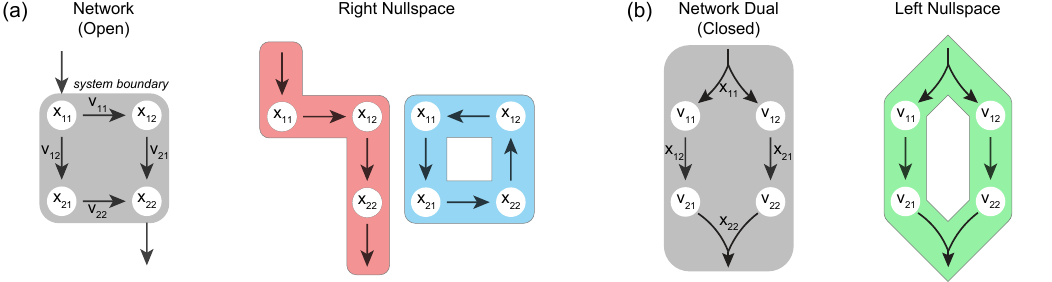}
    \caption{\textbf{The stoichiometric nullspaces of a simple transport network.} Graph nodes $x_{ij}$ and edges $v_{\mu\nu}$ are labeled by their grid position in graph (a) and its dual (b), respectively. The dual (b) of the network (a) is obtained by sending the incidence matrix $U \to -U^\intercal$ as shown in Eq.~(\ref{eq:app-simple-incidence-mat}). The columns and rows of the incidence matrix  are sorted by row-major order of the nodes and edges grid positions. To visualize the nullspace basis elements, we keep the direction of each edge in the graph if the sign of the corresponding basis vector entry is positive and flip if negative. Note that any linear combination of linear subspace basis elements are also elements of the subspace. As such, the right nullspace is a two-dimensional subspace equal to the span of the two basis elements shown in (a).}
    \label{fig:network-example}
\end{figure*}

To develop an intuition about stoichiometric nullspaces, it is useful to consider a basic network architecture with pairwise couplings (edges) between species (nodes) as illustrated in Fig.~\ref {fig:network-example} -- the generalization to more complex networks as in  Fig.~1(b) is then straightforward. In chemistry and biology applications, pairwise-connected networks describe basic conversion processes [Fig.~\ref {fig:network-example}(a)]; in this case, the stoichiometric matrix reduces to the directed incidence matrix $U \in \mathbb{R}^{m\times n}$ of the network. For example, the network in Fig.~\ref{fig:network-example}(a) describes  four internal conversion/transport processes between four species, and the two external edges represent exchange processes with environment. The incidence matrix of the full network is 
\begin{equation}
    U = \begin{bmatrix}
        -1 & -1 &  0 &  0 &  1 &  0 \\
         1 &  0 & -1 &  0 &  0 &  0 \\
         0 &  1 &  0 & -1 &  0 &  0 \\
         0 &  0 &  1 &  1 &  0 & -1 
    \end{bmatrix},
    \label{eq:app-simple-incidence-mat}
\end{equation}
with the last two columns representing external in-flux and out-flux, respectively. The first four columns of $U$ correspond to the incidence matrix of the internal circuit [gray-shaded sub-graph in Fig.~\ref{fig:network-example}(a)], and we denote this submatrix by  
\begin{equation*}
    U_\text{int} = \begin{bmatrix}
        -1 & -1 &  0 &  0 \\
         1 &  0 & -1 &  0 \\
         0 &  1 &  0 & -1 \\
         0 &  0 &  1 &  1 
    \end{bmatrix}.
\end{equation*}
The remaining two columns we collect in the matrix $U_\text{ext}$, such that $U = [U_\text{int} \ U_\text{ext}]$. Of course, more broadly, network  architectures of this type also encompass transport networks, such as electrical circuits~\cite{Stern2022, Stern2025} or pipe systems~\cite{Rocks2019, Alsous2021}, and force networks~\cite{Kane2014, Chen2014, Zhang2022}. In those contexts, the stoichiometric nullspaces have intuitive physical meaning, as briefly discussed next.

\subsubsection{Electrical circuits}

In the language of electrical circuits, chemical species correspond to junctions. We can take the charge at any particular junction as an identifiably distinct species. The temporal evolution of charges at junctions is given by $U i = i_\text{net}$ where $i_\text{net}$ is the vector of net junction currents and $i$ is the vector of currents in the connections. Decomposing $i = [i_\text{int} \ i_\text{ext}]^\intercal$, it follows that 
\begin{equation*}
    U_\text{int} i_\text{int} + U_\text{ext} i_\text{ext} = i_\text{net} .
\end{equation*}
For an open circuit, $i_\text{ext}$ is a vector of externally supplied currents. Taking into account these external currents in the incidence matrix yields the open network of Fig.~\ref{fig:network-example}(a).

\paragraph*{\textbf{Right nullspace.}} 
Notwithstanding any additional driving currents, Kirchhoff's first law requires that the net currents be zero: $Ui = i_\text{net} = 0$. That is, the currents in the connections are elements of the right nullspace of $U$, which is spanned by the columns of 
\begin{equation*}
    \mathcal{N}^{\text{R}}(U) = \mathspan \left\{ \begin{bmatrix}
        1 \\ 0 \\ 1 \\ 0 \\ 1\\ 1
    \end{bmatrix}, \begin{bmatrix}
        -1 \\ 1 \\ -1 \\ 1 \\ 0 \\ 0
    \end{bmatrix} \right\} .
\end{equation*}
We see in Fig.~\ref{fig:network-example}(a) that nonzero entries of these columns correspond to sets of connections in the circuit that when linearly combined satisfy Kirchhoff's first law at every junction. 

\paragraph*{\textbf{Left nullspace.}} 
It is known that the dimension of the left nullspace of an incidence matrix corresponds to the number of connected components of the (closed) graph~\cite{GodsilRoyleGraphTheory}. Our physical intuition of the left nullspace of this matrix is developed by supposing that each of the connections in the electrical circuit has an electrical component with an associated impedance $\hat{z}_{e}$ across connection $e$ between junctions $i$ and $j$. These impedances are defined by the Laplace transform of the voltages across the electrical component and the current through it: 
\begin{equation*}
    \hat{z}_{e}  = \mathcal{L}\{v_{i} - v_{j}\}/\mathcal{L}\{ i_{e} \} = (\hat{v}_{i} - \hat{v}_{j})/\hat{i}_{e} \in \mathbb{C}.
\end{equation*}
This provides Ohm's law in the Laplace-transformed variables which in matrix form is $U^\intercal  \hat{v} = \hat{Z}\hat{i}$ where the vector of complex potentials across connections is $\hat{v} \in \mathbb{C}^n$, $\hat{Z} = \diag(\hat{z}_{e})$ is the diagonal matrix of impedances and $\hat{i} \in \mathbb{C}^n$ is the vector of transformed currents through the connections. 

It follows that in electrical circuits, elements of the left nullspace of the incidence matrix $U$ correspond to a zero voltage differential across electrical components. Namely, for $\hat{\delta} \in \mathcal{N}^\text{L}_\mathbb{C}(U) = \mathcal{N}^\text{L}_\mathbb{R}(U) \otimes_\mathbb{R}\mathbb{C}$, Ohm's law remains unchanged:
\begin{equation*}
    U^\intercal (\hat{v} + \hat{\delta}) = U^\intercal \hat{v} = \hat{Z} \hat{i}.
\end{equation*}
We can then obtain that every element $\hat{\delta}$ of the left nullspace satisfies 
\begin{equation*}
   P(v+\delta) 
   =  ({v}^\intercal +\delta^\intercal) U i 
   =  {v}^\intercal U i = P({v})
\end{equation*}
Namely, the real power contribution of these additional nullspace junction voltages is zero and leads to no physical power exchange. Using that the current is the derivative of charge with respect to time, 
$$
\frac{\text{d} q }{ \text{d} t} \equiv i_\text{net}
$$ 
we also see that elements of the left nullspace correspond to conservation laws for the junction charges: 
\begin{equation*}
 \delta^\intercal \frac{\text{d} q }{\text{d} t} = 0 \quad\implies \quad \delta^\intercal q = \text{const}.
\end{equation*}

For an open network, an element $\delta$ of the left nullspace of $U$ must satisfy 
\begin{equation*} 
\begin{bmatrix}
    U_\text{int}^\intercal \\
    U_\text{ext}^\intercal 
\end{bmatrix} \delta = 
\begin{bmatrix}
    U_\text{int}^\intercal \delta \\
    U_\text{ext}^\intercal \delta
\end{bmatrix} = 0 ,
\end{equation*}
or equivalently,
\begin{equation*}
    U_\text{int}^\intercal \delta = U_\text{ext}^\intercal \delta = 0 .
\end{equation*}
However, with network closure the only constraint $\delta$ must satisfy is $U_\text{int}^\intercal = 0$ from which it follows that the left nullspace of the internal incidence matrix is spanned by the constant vector, $\mathcal{N}^\text{L}(U_\text{int}) = \mathspan \mathbbm{1}$. Physically this means that the absolute voltage level is arbitrary and only voltage differences matter. Thus, the left nullspace of the closed network represents the gauge freedom of the electric potential: the freedom to choose the reference voltage or ground.

In the open network, $U_\text{ext}$ is composed of unit vectors $e_{k}$, the corresponding entries in $\delta$ are fixed as $\delta_k = 0$. Consequently, $U_\text{int}^\intercal \delta = 0$ fixes the remaining entries in $\delta$ to zero. As such the left nullspace of the open network is trivial, $\mathcal{N}^\text{L} (U) = 0$. This means that the voltage reference has been fixed strongly enough so that no nonzero voltage vector is invisible to the circuit.

We can also consider the left nullspace of the incidence matrix in terms of a discrete Laplacian. Imposing Kirchhoff's first law provides that
\begin{equation*}
    \underbrace{U Z^{-1} U^\intercal}_{= \ L} \hat{v} = U \hat{i} = 0 
\end{equation*}
from which we obtain a discrete form of Laplace's equation with Laplacian $L$. For a nonsingular $Z$, the right nullspace of $L$ is the same space as the left nullspace of $U$. Moreover, $L$ is decomposable as the sum of an internal and external Laplacian: 
\begin{equation*}
   L = \underbrace{U_\text{int} Z_\text{int}^{-1} U_\text{int}^\intercal}_{= \ L_\text{int}} + \underbrace{U_\text{ext} Z_\text{ext}^{-1}  U_\text{ext}^\intercal }_{= \ L_\text{ext}}.
\end{equation*} 
We see that an element of the internal Laplacian nullspace must also be an element of the external Laplacian nullspace in order for an element of the left nullspace of the closed network to also be an element of the left nullspace of the open network.

\subsubsection{Mechanical networks}

A closely related interpretation arises for mechanical networks. Suppose that the nodes of the graph represent point masses and the edges represent springs or mechanical constraints between them. The incidence matrix $U$ then maps edge forces to net forces on the nodes. If $f$ is the vector of edge forces and $f_{\mathrm{net}}$ is the vector of external net forces acting on the point masses, then
\begin{equation*}
    U f = f_\text{net}.
\end{equation*}
Mechanical equilibrium requires
\begin{equation*}
    f_{\mathrm{net}}=0,
\end{equation*}
and therefore any balanced force pattern satisfies
\begin{equation*}
Uf=0.
\end{equation*}
Thus, elements of the right nullspace $\mathcal{N}^\text{R}(U)$ correspond to self-balanced force networks: force configurations that generate no net force on any point. This is directly analogous to the role of the right nullspace in chemical reaction networks, where $Sv=0$ defines unconstrained steady-state flux combinations that produce no net accumulation of any chemical species.

\par

The left nullspace has a dual interpretation in terms of displacements. Let $x$ denote the vector of point mass positions. The edge extensions are given by $U^\intercal x$. For a network of linear springs with stiffness matrix $K=\diag(k_e)$, the edge forces satisfy $K U^\intercal x = f$. If $\delta \in \mathcal{N}^\text{L}(U)$, then $U^\intercal \delta=0$, and therefore
\begin{equation*}
    K U^\intercal (x+\delta)=K U^\intercal x=f .
\end{equation*}
Thus, displacements along left nullspace directions do not change the spring extensions and in turn do not change the internal spring forces. Equivalently, these modes have no elastic energy cost. Indeed, the elastic energy is
\begin{equation*}
E(x)=\frac{1}{2}x^\intercal U K U^\intercal x,
\end{equation*}
and for $\delta\in \mathcal{N}^\text{L}(U)$,
\begin{equation*}
    E(x+\delta)=E(x).
\end{equation*}
In this mechanical setting, the left nullspace corresponds to zero-energy displacement modes, such as rigid translations of components or other motions invisible to the network constraints. This mathematical prescription is used in Ref. \cite{Zhang2022} to analyze prestressed systems.

The left nullspace also defines conserved components of momentum. Since Newton’s second law gives
$$f_{\mathrm{net}}=\frac{\dd p_\text{net}}{\dd t},$$
we have, for any $\delta\in \mathcal{N}^\text{L}(U)$,
$$\delta^\intercal f_{\mathrm{net}}
=
\delta^\intercal \frac{\dd p_\text{net}}{\dd t}
=
0,$$
and hence
$$\delta^\intercal p_\text{net}=\mathrm{constant}.$$

As in chemical reaction and electrical networks, the left nullspace of a mechanical network depends on the system boundary. For the closed (internal) network, $U_\text{int}^\intercal \delta=0$ provides displacement modes that do not stretch any internal spring; for a connected network, these include rigid translations. Opening the network by including boundary attachments imposes additional constraints $U_\text{ext}^\intercal \delta=0$. Boundary attachments can therefore remove zero modes that were present in the closed network, just as exchange reactions can break conservation laws in an open chemical reaction network. Although the physical meaning of the left nullspace depends on the system represented by the incidence matrix, these examples illustrate a common principle: left nullspace modes correspond to directions that are invisible to the network dynamics or constraints. In chemical reaction networks, this invisibility appears as conservation of species pools.

\subsection{Conservation laws are broken by opening networks \label{sec:broken-conservation-laws}} 

Here we recapitulate the results of the simple three-component closed chemical reaction network reported in Ref.~\cite{Famili2003}:
\begin{equation*}
    A + B \to C .
\end{equation*}
The corresponding stoichiometric matrix is 
\begin{equation*}
    S_{\text{int}} = \begin{bmatrix}
    -1 \\ -1 \\ 1 
    \end{bmatrix}, 
\end{equation*}
with nullspaces, 
\begin{equation*}
    \mathcal{N}^{\text{R}}(S_{\text{int}}) = 0 \ \ \ \text{and} \ \ \ \mathcal{N}^{\text{L}}(S_{\text{int}}) = 
    \mathspan \left\{  \begin{bmatrix}
       1 \\ 0 \\ 1 
    \end{bmatrix}, 
    \begin{bmatrix}
        0 \\ 1 \\ 1 
    \end{bmatrix} \right \}
\end{equation*}
Note that the right nullspace only consists of the trivial zero vector. The left nullspace provides two conservation laws for the three components: 
\begin{gather*}
    A + C = \text{const} \ \text{and} \ \\ 
    B + C = \text{const}. 
\end{gather*}
If we open the system by allowing exchange of the three components with the environment,
\begin{equation}
    \varnothing \to A, \ \ \  \varnothing \to B, \ \ \ C \to \varnothing ,
    \label{eq:open-system-example}
\end{equation}
the stoichiometric matrix becomes 
\begin{equation*}
    S_{\text{exch}}  = \begin{bmatrix}
    -1 & 1 & 0 & 0 \\ -1 & 0 & 1 & 0 \\ 1 & 0 & 0 & -1 
    \end{bmatrix}
\end{equation*}
where we now obtain a trivial left nullspace and a nontrivial right nullspace: 
\begin{equation*}
    \mathcal{N}^{\text{R}}(S_{\text{exch}}) = \mathspan \begin{bmatrix}
    1 \\ 1 \\ 1 \\ 1
    \end{bmatrix} \ \ \ \text{and} \ \ \ \mathcal{N}^{\text{L}}(S_{\text{exch}}) = 0 . 
\end{equation*}
The nontrivial right nullspace provides that at steady state the fluxes are all equivalent: 
\begin{equation*}
    v_{\text{reaction}} = v_\text{$A$\,influx} = v_\text{$B$\,influx} = v_\text{$C$\,efflux} .  
\end{equation*}

Re-closing the system by replacing Eq.~(\ref{eq:open-system-example}) with 
\begin{equation*}
    A_\text{ext} \to A, \ \ \  B_\text{ext} \to B, \ \ \ C \to C_\text{ext} ,
\end{equation*}
provides the augmented stoichiometric matrix 
\begin{equation*}
S_{\text{tot}} = \begin{bmatrix}
    -1 & 1 & 0 & 0 \\ 
    -1 & 0 & 1 & 0 \\ 
    1 & 0 & 0 & -1 \\
    0 & -1 & 0 & 0 \\
    0 & 0 & -1 & 0 \\
    0 & 0 & 0 & 1
    \end{bmatrix} ,
\end{equation*}
where we again obtain a trivial right nullspace and nontrivial left nullspace: 
\begin{equation*}
    \mathcal{N}^{\text{R}}(S_{\text{tot}}) = 0  \ \ \ \text{and} \ \ \ \mathcal{N}^{\text{L}}(S_{\text{tot}}) = \mathspan \left\{ 
    \begin{bmatrix}
    1 \\ 0 \\  1 \\ 1 \\  0 \\  1
    \end{bmatrix},
    \begin{bmatrix}
    0 \\ 1 \\  1 \\  0 \\ 1 \\ 1
    \end{bmatrix}
    \right\}. 
\end{equation*}
The total concentration of the components are then constrained by the conservation laws
\begin{gather*}
    A_{\text{tot}} + C_{\text{tot}} = \text{const} \ \text{and} \\
    B_{\text{tot}} + C_{\text{tot}} = \text{const},
\end{gather*}
where $X_{\text{tot}} = X + X_{\text{ext}}$ for $X \in \{A,B,C\}$. 

These examples show that the left nullspace is not an intrinsic property of the internal reaction list alone; it depends on the system boundary. For this reason, open- and closed-network left nullspaces answer different physical questions. Therefore, in the main text, we primarily compute the Grassmann distances from the same open-network stoichiometric matrix, while closed-network left nullspace distances are used only as a probe of internal conservation structure. 

\subsection{Rescaling stoichiometric coefficients}

Column-wise rescaling the stoichiometric matrix according to Eq.~(\ref{eq:REACTION-RESCALING}) leaves the chemistry of the chemical reaction network unchanged as the physically relevant ratios are preserved. The right nullspace is inherently sensitive to such rescalings. We prove that this leads to changes in the R-Grassmann distance which are bounded by the largest and smallest entries in the scaling matrix $\Lambda$. Empirically, we find that most changes in mini-AGORA2 R-Grassmann distances from natural choices of $\Lambda$ are small relative to their unscaled counterparts. 

Let two subspaces of equal dimension $X$ and $Y$ be the endpoints of a path $\gamma(t)$ in the infinite Grassmannian $\Gr(p,\infty)$. Let $U(t)$ be a particular choice of orthogonal frame representation. Consequently, 
\begin{equation*}
    U(t)^\intercal U(t) = \mathbbm{1} ,
\end{equation*}
from which it follows that 
\begin{equation*}
    \frac{\text{d}}{\text{d}t} [U^\intercal U ] 
    = \dot{U}^\intercal U + 
    U^\intercal \dot{U} = 0 .
\end{equation*}
$U^\intercal \dot{U}$ is therefore skew-symmetric. 

We impose that $U(t)$ obeys the horizontal-frame condition, $U^\intercal \dot{U} = 0$. This is a choice of gauge that removes the within-subspace component of the frame's motion (Section 2.5 in Ref. \cite{Edelman1998}). Namely, the frame has no instantaneous rotation within its own column space and can always be arranged along a smooth path. To make this apparent consider the following. Suppose that we have an alternate choice of frame $V(t)$ and define
\begin{equation*}
    A(t) = V^\intercal \dot{V}, 
\end{equation*}
which, like $U^\intercal \dot{U}$, is skew-symmetric. Using that the Lie algebra of the orthogonal matrices are the skew-symmetric matrices, we can then choose an orthogonal matrix $O(t)$ that satisfies 
\begin{equation*}
    \dot{O}(t) = -A(t) O(t) 
\end{equation*}
and define 
\begin{equation*}
U(t) = V(t) O(t).
\end{equation*}
Importantly, right-multiplication by $O(t)$ does not alter the represented subspace; only the choice of frame within the column space of the subspace is changed. It follows that 
\begin{align*} 
U^\intercal \dot{U} &= O^\intercal V^\intercal (\dot{V} O +  V\dot{O} )\\
&=O^\intercal A O + 
O^\intercal \dot{O} \\
&=O^\intercal A O -
O^\intercal AO = 0 .
\end{align*}

For a general orthogonal frame, the Grassmannian speed of the path is equal to the magnitude of the velocity component orthogonal to the subspace:
\begin{equation*}
|\gamma'(t)| = 
\lVert (I - UU^\intercal ) \dot{U}\rVert_F,
\end{equation*} 
which under the horizontal-frame condition reduces to $\lVert \dot{U}\rVert_F$. 

Now if we transform the path $\gamma(t)$ according to a fixed invertible matrix $M$, the resulting path $\tilde\gamma(t) = M\gamma(t)$ has a transformed orthogonal frame representation $Q(t)$ satisfies $M U(t) = Q(t) R(t) $ where $R(t)$ is invertible and upper triangular, exactly the thin QR decomposition of $MU(t)$. Rearranging provides
\begin{equation*}
Q(t) = M U(t) R^{-1}(t).
\end{equation*} 
Observe that $R^\intercal R = R^\intercal Q^\intercal Q R = U^\intercal M^\intercal M U$. Let us define the symmetric matrix
\begin{equation*}
    H(t) = U^\intercal(t) M ^\intercal M U(t) = R^\intercal (t) R (t). 
\end{equation*}
Since $H= R^\intercal R$, $H^{-1} = R^{-1} R^{-1\intercal}$ where $R^{-1}$ is also upper triangular. Clearly, $R^{-1}$ behaves as a Cholesky-like factor of $H^{-1}$. Therefore, any matrix square root satisfying $(H^{-1/2})(H^{-1/2})^\intercal = H^{-1}$ must be of the form $H^{-1/2} O = R^{-1}$ where $O$ is an orthogonal matrix. Therefore, dropping the right orthogonal factor which does not alter the represented subspace, we obtain
\begin{equation*}
Q(t) = M U(t) H^{-1/2}(t).
\end{equation*} 
which is like the original thin QR-decompositiond. Now we would like to identify the Grassmannian speed of the orthogonal frame $Q(t)$
\begin{equation*}
|\tilde\gamma'(t)| = 
\lVert (I - QQ^\intercal ) \dot{Q}\rVert_F.
\end{equation*} 
Differentiating $Q(t)$ gives 
\begin{equation*}
\dot{Q}(t) = M \dot{U} H^{-1/2} + M U \frac{\text{d}}{\text{d}t}H^{-1/2}.
\end{equation*} 

Since $MU = Q H^{1/2}$
\begin{equation*}
M U \frac{\text{d}}{\text{d}t}H^{-1/2}
= Q \underbrace{H^{1/2} \frac{\text{d}}{\text{d}t}H^{-1/2}}_{= K},
\end{equation*}
and using that the projection operator $I - Q Q^\intercal$ annihilates terms of the form $QK$,
\begin{equation*}
(I - QQ^\intercal ) QK  = QK - QQ^\intercal Q K =0,
\end{equation*} 
the Grassmannian speed reduces to
\begin{equation*}
|\tilde\gamma(t)|  =  \lVert (I - QQ^\intercal )  M \dot{U} H^{-1/2} \rVert_F.
\end{equation*} 
Orthogonal projection cannot increase the Frobenius norm, therefore, 
\begin{equation*}
|\tilde\gamma(t)|  \leq \lVert  M \dot{U} H^{-1/2} \rVert_F,
\end{equation*} 
Spectral bounds on the Frobenius norm give 
\begin{equation*}
|\tilde\gamma'(t)|
\leq \lVert  M \rVert_2 \underbrace{\lVert \dot{U} \rVert_F }_{= \ |\gamma'(t)|}
\lVert H^{-1/2} \rVert_2,
\end{equation*} 
where $\lVert \cdot  \rVert_2 $ is the operator 2-norm and is equal to the largest singular value of the argument, $\sigma_\text{max}(\cdot)$.

What can we say about $\lVert H^{-1/2} \rVert_2$? For any $z \in \mathbb{R}^p$, 
\begin{equation*}
z^\intercal H z = z^\intercal U^\intercal M^\intercal M U z = \lVert M U z \rVert_2^2,
\end{equation*} 
The smallest singular value of $M$ satisfies 
\begin{equation*}
\lVert M y \rVert_2 \geq \sigma_\text{min}(M) \lVert y \rVert_2.
\end{equation*}
Using $y = Uz$ and $U^\intercal U = \mathbbm{1}$, 
\begin{equation*}
z^\intercal H z = \lVert M U z \rVert_2^2
\geq \sigma_\text{min}(M)^2 \lVert Uz \rVert_2^2
= \sigma_\text{min}(M)^2 \lVert z \rVert_2^2.
\end{equation*}
This Rayleigh-quotient says that every eigenvalue of the symmetric $H$ is at least $\sigma_\text{min}(M)^2$. Therefore, 
\begin{equation*}
    \lVert H^{-1/2} \rVert_2 = \frac{1}{\sqrt{\lambda_\text{min}(H)}} \leq \frac{1}{\sigma_\text{min}(M)}
\end{equation*}
and consequently
\begin{equation*}
|\tilde\gamma'(t)| 
\leq \frac{ \sigma_\text{max}(M)}{\sigma_\text{min}(M) }|\gamma'(t)|  .
\end{equation*} 
Integrating along paths preserves the inequality for path lengths,
\begin{equation*}
    L[\tilde\gamma]\leq \frac{ \sigma_\text{max}(M)}{\sigma_\text{min}(M) }\, L[\gamma]
\end{equation*}
The extension to geodesics on the Grassmannian \cite{Edelman1998,Absil2004} is straightforward: the geodesic distance of the image of the curve under $M$ between $MX$ and $MY$ cannot exceed the length of the image curve. Therefore,
\begin{equation*}
    \dist(MX,MY) \leq L[\tilde\gamma] \leq \frac{ \sigma_\text{max}(M)}{\sigma_\text{min}(M) }\, L[\gamma] .
\end{equation*}
As this is valid for every $\gamma$, we may take the infimum over all paths $\gamma$:
\begin{align*}
    \dist(MX,MY) &\leq \inf_\gamma\left(\frac{ \sigma_\text{max}(M)}{\sigma_\text{min}(M) }\, L[\gamma] \right) \\
    &= \frac{ \sigma_\text{max}(M)}{\sigma_\text{min}(M) } \inf_\gamma L[\gamma] \\
    &= \frac{ \sigma_\text{max}(M)}{\sigma_\text{min}(M) }\dist(X,Y).
\end{align*}
Thus we ultimately arrive at
\begin{equation*}
    \dist(M X, MY) \leq \cond(M) \dist(X, Y)
\end{equation*}
where we define the condition number $\cond(M) = \sigma_\text{max}(M)/\sigma_\text{min}(M)$. Moreover, 
\begin{align*}
    \dist(M^{-1}M X, M^{-1}MY) &= \dist(X, Y) \\
    & \leq \cond(M^{-1}) \dist(MX, MY) \\
    &= \cond(M) \dist(MX, MY),
\end{align*}
where we use that the condition number of an invertible matrix $M$ does not change due to matrix inversion. Thus, 
\begin{align*}
     \frac{1}{\cond(M)} \dist(X, Y)
     &\leq \dist(M X, MY)\\
     &\leq \cond(M) \dist(X, Y).
\end{align*}

Now, if $X$ and $Y$ are not necessarily subspaces of equal dimension, we must work with the common-space distance $\delta$. If $X$ and $Y$ are $k$- and $\ell$-dimensional subspaces, where $k \leq \ell$, 
\begin{align*}
    \delta(X,Y) &= \min_{T\in \Omega_-(Y)} \dist(A,T),
\end{align*}
where $\Omega_-(Y)$ is the Schubert variety of $Y$ which contains the $k$-dimensional subspaces contained in $Y$ \cite{Ye2016}. Moreover, because $M$ is invertible $T \mapsto MT$ is a bijection from $\Omega_-(Y)$ to $\Omega_-(MY)$. Therefore, 
\begin{align*}
    \delta(MX,MY) &= \min_{S\in \Omega_-(MY)} \dist(MX,S) \\
    &= \min_{T\in \Omega_-(Y)} \dist(MX,MT).
\end{align*}
Taking minima over the candidate set $\Omega_-(Y)$ yields, 
\begin{align*}
    &\frac{1}{\cond(M)} \min_{T\in \Omega_-(Y)} \dist(X, T) = \frac{1}{\cond(M)} \delta(X,Y) \\
    & \hspace{10ex} \leq \min_{T\in \Omega_-(Y)} \dist(M X, MT) = \delta(MX,MY) \\
    & \hspace{10ex} \leq  \cond(M) \min_{T\in \Omega_-(Y)}  \dist(X, T) \\
    & \hspace{10ex} = \cond(M) \delta(X,Y).
\end{align*}
The same holds for $k >\ell$, after taking minima over the candidate set $\Omega_-(X)$. Therefore,
\begin{equation*}
     \frac{1}{\cond(M)} \delta(X, Y)
     \leq \delta(M X, MY) \leq \cond(M) \delta(X, Y),
\end{equation*}
from which we then find that 
\begin{align*}
     & \left[ |k - \ell|\frac{\pi^2}{4} + \frac{1}{\cond(M)^2} \delta(X, Y)^2 \right]^{1/2}
     \leq  \dist(M X, MY) \\
     & \hspace{10ex} \leq \left[ |k - \ell|\frac{\pi^2}{4} +  \cond(M)^2 \delta(X, Y)^2 \right]^{1/2}.
\end{align*}

$\dist(MX, MY)$ is almost exactly the distance we would like to bound:
\begin{equation*}
    \dist(M_X X, R M_X Y)
\end{equation*}
where $R = M_Y M_X^{-1}$. While the results above hold for any given invertible $M$, we now need to require that $R$ is an invertible symmetric positive definite matrix. This clearly holds if $M_X$ and $M_Y$ are diagonal matrices with positive entries.  

Consider $v \in M_X Y$ where $|v| = 1$. We would like to bound the change in $v$ due to $R$. We may always decompose $Rv$ in terms of two components, one parallel to $v$ and one orthogonal to $v$:
\begin{equation*}
    Rv = (v^\intercal Rv) v + (R v - (v^\intercal Rv)v). 
\end{equation*}
Now the component that is orthogonal to $v$ has magnitude 
\begin{equation*}
    v^\intercal R^\intercal R v - (v^\intercal Rv)^2 = v^\intercal R^2 v - \mu^2,
\end{equation*}
where we define $v^\intercal Rv = \mu$. It follows that the angle between $v$ and $Rv$ satisfies 
\begin{equation*}
    \tan^2 \angle(v , Rv) = \frac{v^\intercal R^2 v - \mu^2}{\mu^2}.
\end{equation*}
Every eigenvalue of $R$, $\lambda_k\in \lambda(R)$ is in the interval $[\lambda_\text{min}(R),\lambda_\text{max}(R)]$. Therefore, 
\begin{gather*}
    (\lambda_k - \lambda_\text{min})(\lambda_k - \lambda_\text{max}) \leq 0 \\ 
    \implies 
    \lambda_k^2 \leq (\lambda_\text{min}+\lambda_\text{max})\lambda_k - \lambda_\text{min}\lambda_\text{max}.
\end{gather*}
Expanding $v$ in an orthonormal eigenbasis of $R$:
\begin{equation*}
    v = \sum_k a_k e_k \ \ \ \text{where} \ \sum_k a_k^2 = 1,
\end{equation*}
we observe that
\begin{align*}
v^\intercal R v &=  \sum_{k,\ell} a_\ell a_k e_\ell^\intercal R e_k 
    =  \sum_{k,\ell} \lambda_k a_\ell a_k \delta_{k\ell} = \sum_{k} \lambda_k a_k^2\\
    v^\intercal R^2 v &=  \sum_{k,\ell} a_\ell a_k e_\ell^\intercal R^2 e_k 
    =  \sum_{k,\ell} \lambda_k^2 a_\ell a_k \delta_{k\ell} = \sum_{k} \lambda_k^2 a_k^2,
\end{align*}
from which it follows that 
\begin{equation*}
   v^\intercal R^2 v \leq (\lambda_\text{min} + \lambda_\text{max})v^\intercal R v - \lambda_\text{min}\lambda_\text{max} .
\end{equation*}
Therefore,
\begin{align*}
    \tan^2 \angle(v, Rv ) &\leq \frac{(\lambda_\text{min} + \lambda_\text{max}) \mu - \lambda_\text{min}\lambda_\text{max} - \mu^2  }{\mu^2}  \\
    &= \frac{(\lambda_\text{max}-\mu)(\mu-\lambda_\text{min})}{\mu^2}.
\end{align*}
Now $\mu$ also exists in the interval $[\lambda_\text{min},\lambda_\text{max}]$. This particular upper bound on $\tan^2\angle (v, Rv)$ is maximized by $\mu_*$ satisfying 
\begin{equation*}
    \mu_* = 2\frac{\lambda_\text{min} \lambda_\text{max}}{\lambda_\text{min} + \lambda_\text{max}}.
\end{equation*}
Therefore, 
\begin{align*}
    \tan^2 \angle(v, Rv ) &\leq 
    \left( \frac{\lambda_\text{min} + \lambda_\text{max}}{2 \lambda_\text{min}} - 1 \right)   \left( 1 - \frac{\lambda_\text{min} + \lambda_\text{max}}{2 \lambda_\text{max}}\right) \\
    &=  \frac{(\lambda_\text{max} -\lambda_\text{min})^2}{4 \lambda_\text{min}  \lambda_\text{max}}  .
\end{align*}
Defining the eigenvalue condition number as $\cond_\lambda(R) = \lambda_\text{max}(R)/\lambda_\text{min}(R)$, 
\begin{equation*}
\angle(v, Rv ) \leq \arctan
    \frac{\cond_\lambda(R) - 1}{2 \sqrt{\cond_\lambda(R) }}  .
\end{equation*}
Moreover, using that $\arctan x = \arcsin (x/\sqrt{1+x^2})$,
\begin{equation*}
\angle(v, Rv ) \leq \arcsin
    \frac{\cond_\lambda(R) - 1}{\cond_\lambda(R) + 1} \equiv \alpha(R) .
\end{equation*}
Now if $T$ is a $p$-dimensional subspace, all principal angles must be at most $\alpha(R)$ from which we bound the Grassmann distance between the subspace and its image under $R$ by
\begin{equation*}
    \dist(T,RT) \leq \sqrt{p} \, \alpha(R).
\end{equation*}

Let us consider the case where $A$ and $B$ are $k$- and $\ell$-dimensional subspaces, respectively. If $k\leq \ell$
\begin{align*}
    \delta(A,B) &= \min_{T\in \Omega_-(B)} \dist(A, T) \\ 
    \delta(A,RB) &= \min_{S\in \Omega_-(RB)} \dist(A,S) = \min_{T\in \Omega_-(B)}\dist(A,RT)
\end{align*}
The triangle inequality provides that
\begin{equation*}
    |\dist(A,RT) - \dist(A,T)| \leq \dist(T,RT) \leq \sqrt{k} \, \alpha(R).
\end{equation*}
Taking minima over the candidate set $\Omega_-(B)$ ultimately yields 
\begin{equation*}
    |\delta(A,RB) - \delta(A,B)| \leq \sqrt{k} \, \alpha(R).
\end{equation*}
Analogously, if we instead have $k > \ell$,
\begin{align*}
    \delta(A,B) &= \min_{T\in \Omega_-(A)} \dist(T, B)\\
    \delta(A,RB) &= \min_{T\in \Omega_-(A)} \dist(T, RB), 
\end{align*}
which means that by the triangle inequality,
\begin{equation*}
    |\dist(T,RB) - \dist(T,B)| \leq \dist(B,RB) \leq \sqrt{\ell} \, \alpha(R),
\end{equation*}
such that by taking the minima over the candidate set $\Omega_-(A)$,
\begin{equation*}
    |\delta(A,RB) - \delta(A,B)| \leq \sqrt{\ell} \, \alpha(R).
\end{equation*}
Thus we arrive at the result that 
\begin{equation*}
    |\delta(\Lambda_X X,R\Lambda_XY) - \delta(\Lambda_X X, \Lambda_X Y)| \leq \underbrace{\sqrt{\min(k,\ell)}\,\alpha(R)}_{= \, \varepsilon},
\end{equation*}
where we define $\varepsilon = \sqrt{\min(k,\ell)}\,\alpha(R)$ is the correction due $R = \Lambda_Y \Lambda_X^{-1}$.

Combining both bounds on the common-space distance we arrive at
\begin{align*}
    &\max\{\delta(\Lambda_X X, \Lambda_X Y) - \varepsilon,0\} \leq \delta(\Lambda_X X,\Lambda_YY) \\
    &\hspace{10ex}\leq \min\{\delta(\Lambda_X X, \Lambda_X Y) + \varepsilon, \sqrt{\min(k,\ell)}\pi/2 \}
\end{align*}
and 
\begin{equation*}
    \frac{1}{\cond(\Lambda_X)} \delta(X, Y)
     \leq \delta(\Lambda_X X, \Lambda_X Y) \leq \cond(\Lambda_X) \delta(X, Y).
\end{equation*}
Moreover, if we had instead initially transformed according to $\Lambda_Y$ we arrive at 
\begin{align*}
    &\max\{\delta(\Lambda_Y X, \Lambda_Y Y) - \varepsilon,0\} \leq \delta(\Lambda_X X,\Lambda_YY)  \\
    & \hspace{10ex}\leq \min\{\delta(\Lambda_Y X, \Lambda_Y Y) + \varepsilon, \sqrt{\min(k,\ell)}\pi/2\} 
\end{align*}
\begin{equation*}
     \frac{1}{\cond(\Lambda_Y)} \delta(X, Y)
     \leq \delta(\Lambda_Y X, \Lambda_Y Y) \leq \cond(\Lambda_Y) \delta(X, Y).
\end{equation*}
Defining $\kappa_* = \min\{\cond(\Lambda_X),\cond(\Lambda_Y)\}$, we combine both sets of inequalities to obtain 
\begin{align*}
    &\left[\frac{1}{\kappa_*} \delta(X, Y) - \varepsilon\right]_+  \leq \delta(\Lambda_X X,\Lambda_YY) \\
    & \hspace{5ex} \leq \min\left\{\kappa_* \delta(X,Y) + \varepsilon, \sqrt{\min(k,\ell)}\pi/2 \right\}
\end{align*}
where we use the notation $[z]_+ = \max(z,0)$. Thus, for the full distance we finally arrive at 
\begin{gather*}
    \underbrace{\left[ \left[\frac{1}{\kappa_*} \delta(X, Y) - \varepsilon\right]_+^2 + |k-\ell| \frac{\pi^2}{4} \right]^{1/2}}_{=\, L} \leq \dist(\Lambda_X X,\Lambda_YY) \\
    \leq \underbrace{\left[  \min\left\{\kappa_* \delta(X,Y) + \varepsilon, \sqrt{\min(k,\ell)}\pi/2 \right\}^2 + |k-\ell| \frac{\pi^2}{4} \right]^{1/2}}_{= \ U}.
\end{gather*}

Defining the lower and upper bound as $L$ and $U$, respectively,
\begin{align}
    &|\dist(\Lambda_X X,\Lambda_YY) - \dist(X,Y) |\nonumber \\
   &\hspace{5ex} \leq \max\{U - \dist(X,Y),  \dist(X,Y) - L\}
    \label{eq:RESCALING-BOUND}
\end{align}
Thus, rescaling the stoichiometric matrices by Eq.~(\ref{eq:REACTION-RESCALING}) will never lead to an unbounded change in the Grassmann distance between right nullspaces. Critically, the upper bound dependence on $\Lambda$ is only from the smallest and largest diagonal entries.  

\begin{figure}
\centering
\includegraphics[width=8.7cm]{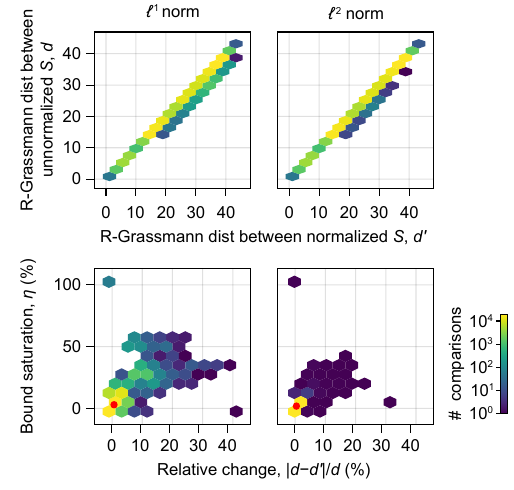}
    \caption{\textbf{Column-wise normalization of the stoichiometric matrix leads to bounded changes in R-Grassmann distance between mini-AGORA2 networks.} (Top) $\ell^1$ and $\ell^2$ column-wise normalization of mini-AGORA2 stoichiometric matrices leads to changes in R-Grassmann distances, $d$ versus $d'$, that are (Bottom) empirically bounded in relative change $|d-d'|/d$ and less than $\sim$10\% of the upper bound shown in Eq.~(\ref{eq:RESCALING-BOUND}). 100\% bound saturation is realized by a handful of distances due to zero relative change and upper bound. The median of each statistic is shown in red: most distances attain bound saturation are $ < 3\%$ and relative change $< 1\%$, indicating that most empirical R-Grassmann distances are robust to reasonable choices of column-wise normalization, while also not saturating the upper bound in Eq.~(\ref{eq:RESCALING-BOUND}). 
    \label{fig:RNS-normalization}}
\end{figure}

We now consider two particular choices for $\Lambda$. 
Let $S = [\col_1(S),\ldots, \col_n(S)]$ where $\col_i$ is the $i$-th column of $S$ and 
\begin{equation*}
    \Lambda_p = \diag(1/\lVert \col_1(S) \rVert_p,\ldots, 1/\lVert\col_n(S)\rVert_p).
\end{equation*}
The product $S\Lambda_p$ normalizes the columns of $S$ by their $\ell^p$-norms. $p=1$ is a natural choice as the the $\ell^1$-norm then accounts for the number of particles involved in each metabolic process. $p = 2$ provides an alternative and mathematically motivated normalization choice as the $\ell^2$-norm is related to the standard inner product of Euclidean space. 

Let us define the saturation ratio 
\begin{equation*}
    \eta \equiv \frac{|\dist(\Lambda_X X,\Lambda_YY) - \dist(X,Y) |}{\max\{U - \dist(X,Y), \dist(X,Y)-L\}) 
    }.
\end{equation*}
If $|\dist(\Lambda_X X,\Lambda_YY) - \dist(X,Y) | = \max\{U - \dist(X,Y), \dist(X,Y) - L\}  = 0$, we will take $\eta = 1$. For the mini-AGORA2 dataset, we find that most networks realize $\eta < 3\%$ across all pairwise comparisons of unique networks where pairs of unnormalized stoichiometric matrices are compared against their $\ell^1$- or $\ell^2$-normalized counterparts, with 100\% saturation due to an upper bound of zero [Fig.~\ref{fig:RNS-normalization}]. Therefore, the upper bound is not empirically saturated. Moreover, most relative changes in distance are less than $1 \%$ despite a relatively small number of distances changing by at most $\approx42\%$, indicating that most empirical R-Grassmann distances are robust to $\ell^1$- and $\ell^2$-normalization. While R-Grassmann distances change, we find that the strongest phenotype associations persist under these reasonable column-wise normalizations, although effect sizes and statistical significance can vary (Supplemental Table 2).

\section{GRASSMANN DISTANCE AND NETWORK SIZE\label{sec:dist-components}}

\subsection{Rank-nullity imposes constraints on dimension size}
Suppose that we have stoichiometric matrix $S \in \mathbb{R}^{m\times n}$. The rank-nullity theorem asserts that the dimension of its right nullspace is 
\begin{equation*}
    \dim \mathcal{N}^\text{R}(S) = n - \rank S. 
\end{equation*}
The stoichiometric matrices of open metabolic networks generally have more columns than rows, $m < n$ [Fig.~\ref{fig:size-nullity}(a)]. Therefore, the rank of $S$ is bounded from above by the number of rows it contains since $\rank S \leq \min(m,n)$. Taking $\rank S = \rho m$, 
\begin{equation*}
    \dim \mathcal{N}^\text{R}(S) = n - \rho m,
\end{equation*} 
where $\rho \in [0,1]$ is the ratio of $\rank S$ to its theoretic maximum, $m$. The analogous statement for the left nullspace is obtained by sending $S\to S^\intercal$ and noting that $\rank S = \rank S^\intercal$:
\begin{equation*}
    \dim \mathcal{N}^\text{L}(S) = (1-\rho)m. 
\end{equation*}
Sending $\rho \to 0$ provides the low-rank limit that the dimensions of the right and left nullspaces are equal to the number of columns and rows, respectively, and otherwise bounded from above by these quantities. Conversely, $\rho \to 1$ gives the full-rank limit that the dimension of the right nullspace is equal to the difference between the number of columns and rows, whereas the dimension of the left nullspace must be exactly zero. Similar statements can be made for $S$ with $m \geq n$. Hence, for matrices of the same size, the dimensions of the nullspaces are determined by the value of $\rho$ which then goes into the calculation of the dimension gap in the Grassmann distance. 

Thus, the dimension-gap contribution to the Grassmann distance partly reflects differences in the number of metabolic processes, species, and stoichiometric constraints, whereas the angular term captures differences in the orientation of the nullspaces after accounting for these dimensions.

\begin{figure*}[bt!]
\centering
\includegraphics[width=17.8cm]{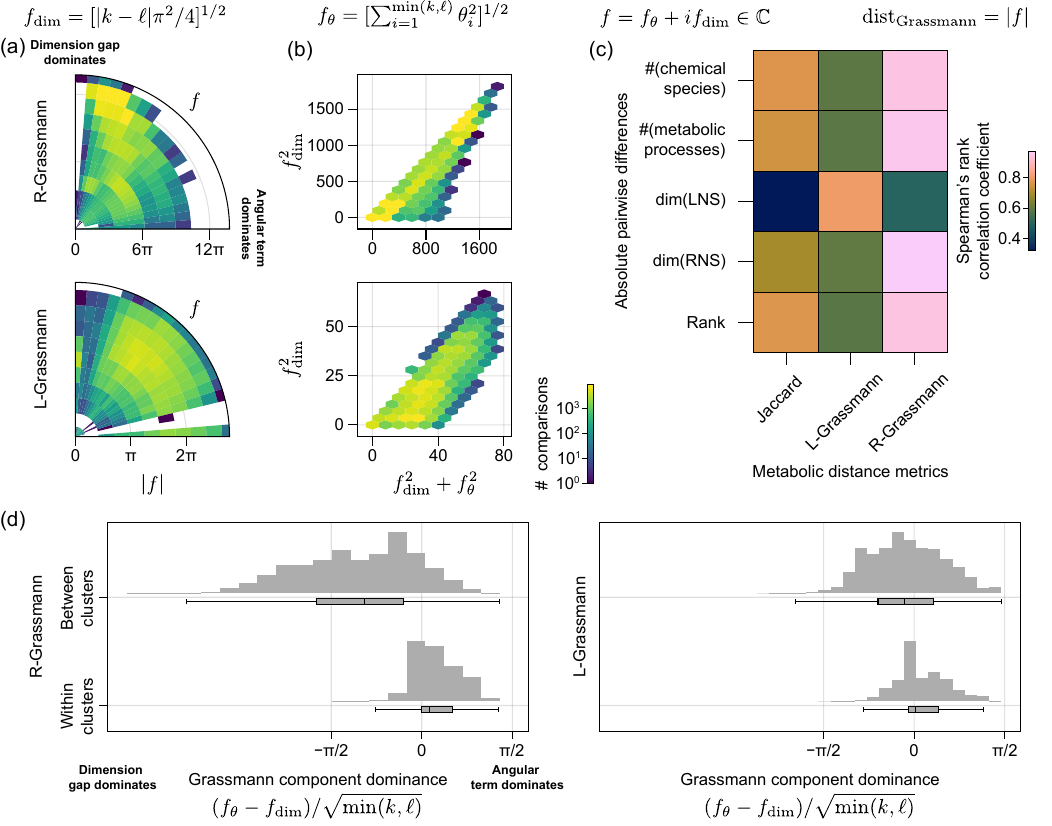}
    \caption{\textbf{Grassmann distances capture more than network size.} 
    (a) Pairwise mini-AGORA2 Grassmann distances are decomposed into dimension gap and angular term components according to Eq.~(\ref{eq:COMPLEX-DECOMPOSITION-GRASSMANN}). Components are well-distributed in the nonnegative quadrant of the complex plane indicating that distances are not primarily dominated by either component.
    As expected, (b) scatter plots and (c) Spearman correlation between pairwise distances and pairwise network-summary differences show that the Grassmann distances are correlated to differences in network size and nullity. (d) mini-AGORA2 distance matrices are separated into within- and between-cluster submatrices. In this manner, block-diagonal entries correspond to within cluster comparisons, whereas off-diagonal entries correspond to between cluster comparisons. Between-cluster separations are more strongly influenced by dimension gap which is lost when considering within-cluster distances, indicating that the clusters reflect both nullity differences and subspace orientation. Box-and-whisker plots show the quartiles with error bars corresponding to 1.5$\times$IQR beyond Q1 and Q3. 
    } \label{fig:size-effects} 
\end{figure*}

\subsection{The Grassmann distance captures more than network size}

The Grassmann distance metric [Eq.~(\ref{eq:GRASSMANN})] can be recast as the modulus of a complex number $f$ with components
\begin{equation}
    f = f_\theta + i f_\text{dim} 
    \label{eq:COMPLEX-DECOMPOSITION-GRASSMANN}
\end{equation}
where the real and imaginary parts,
\begin{equation*}
\textstyle 
f_\theta = [\sum_{i=1}^{\min(k,\ell)} \theta_i^2]^{1/2}
\ \ \ \text{and} \ \ \ 
f_\text{dim} = [|k-\ell|\pi^2/4]^{1/2} ,
\end{equation*} 
capture the angular term or common-space distance \cite{Ye2016} and dimension gap, respectively. Grassmann distances can then be plotted in the 2D plane. If a Grassmann distance is dominated by its dimension gap, it will be placed closer to the imaginary axis, whereas a distance dominated by the angular term is placed closer to the real axis. We find that for both stoichiometric nullspaces of the mini-AGORA2 dataset, pairwise distances are well-distributed in the nonnegative quadrant of the complex plane [Fig.~\ref{fig:size-effects}(a)]. Moreover, true to form, the squared modulus is linearly related to the difference in nullity, but this relationship does not account for the entirety of the variance observed in the Grassmann distance [Fig.~\ref{fig:size-effects}(b)]. For example, the R-Grassmann distance between pairs (\textit{Aggregatibacter aphrophilus} NJ8700, \textit{Alistipes shahii} WAL 8301) and (\textit{Alistipes shahii} WAL 8301, \textit{Bacteroides dorei} DSM 17855) have equivalent $f_\text{dim} = 0$ but unequal $f_\theta$, $\approx 41.48$ versus $\approx 50.60$, respectively. For the L-Grassmann distance, we have pairs (\textit{Actinomyces odontolyticus} ATCC 17982, \textit{Aeromonas hydrophila} 116) and (\textit{Actinomyces odontolyticus} ATCC 17982, \textit{Yokenella regensburgei} ATCC 43003) with $f_\text{dim} \approx 2.47$, with $f_\theta$ differing by an order of magnitude, $\approx2.54$ versus $\approx44.48$, respectively. Therefore, using only the difference in nullity would neglect information, whereas the full Grassmann distance allows us to further discriminate between different metabolic networks. 

As expected from rank-nullity, Spearman's rank correlation coefficient indicates that the R-Grassmann distance strongly correlates with differences in network size (number of processes and species), rank, and nullity, whereas the L-Grassmann distance does so to a lesser extent [Fig.~\ref{fig:size-effects}(c)]. These correlations are not limitations of the metrics; the dimensions of the nullspaces are a stoichiometric property of the network. However, the phenotype-prediction [Fig.~\ref{fig:phenotype-prediction}] and angular contribution [Fig.~\ref{fig:size-effects}] analyses indicate that Grassmann distances also capture variation in subspace orientation beyond size differences. 

The correlation between the Jaccard distance and network size is similarly expected [Fig.~\ref{fig:size-effects}(c)]. While the Jaccard distance is not primarily a measure of set size difference, but rather uniqueness, in the case of set subsumption it is proportional to the magnitude of the set size difference. This is evident in the case of Eq.~(\ref{eq:KO-JACCARD}) and made explicit by the following decomposition. Suppose that two sets $A$ and $B$ are partitioned such that 
\begin{equation*}
    A = A^* \cup S \ \ \ \text{and} \ \ \ 
    B = B^* \cup S
\end{equation*}
where $S = A \cap B$ and $A^*\cap S = B^*\cap S  = \varnothing$. Then the cardinality of the sets is given by $\#(A) = \#(A^*) + \#(S)$ and $\#(B) = \#(B^*) + \#(S)$. It follows that the Jaccard distance between these two sets is 
\begin{align*}
    \dist_\text{Jaccard}(A,B)&= 1 - 
    \frac{\#(A\cap B)}{\#(A \cup B)} \\
    &= 1 - \frac{\#(S)}{\#(A^*) + \#(B^*) + \#(S)}  \\
    &= \frac{\#(A^*) + \#(B^*)}{\#(A^*) + \#(B^*) + \#(S)} 
\end{align*}
where 
$\#(A^*) + \#(B^*) \in [|\#(A) - \#(B)|, \#(A) + \#(B) ]$. In the regime where either set subsumes the other, the lower bound is realized. Thus, in this case, the Jaccard distance is proportional to the magnitude of the set size difference. 

\paragraph*{\textbf{Intra- and inter-cluster separation are associated with different Grassmann components.}} For the eight clusters shown in Fig.~\ref{fig:metabolic-clusterings}, we decompose each pairwise distance into $f_\theta$ and $f_\text{dim}$, and compared the contribution of these terms within- and between-clusters by computing the Grassmann component dominance,
\begin{equation*}
   \Delta f = \frac{f_\theta - f_\text{dim}}{\sqrt{\min(k,\ell)}}.
\end{equation*}
The blockwise analysis shows that the Grassmann clusters are not solely nullity-based: while between-cluster distances are mostly due to the dimension gap, this dominance is lost when examining within-cluster structure which displays more angular contribution [Fig.~\ref{fig:size-effects}(d)]. 

\section{METABOLIC DISTANCES OVER CHEMOTONS\label{sec:chemoton}}

Ganti's chemoton~\cite{Ganti1975,Csendes1984} provides a minimal analytically tractable model for clarifying how nullspace-based distances differ from both set-overlap and flux vector comparisons. Because stoichiometric coefficients and nutrient influxes can be varied independently, the chemoton model exposes regimes in which Jaccard distances are insensitive or degenerate and regimes in which distances between flux balance analysis (FBA) solutions depend explicitly on environmental assumptions and optimization choices. In contrast, Grassmann distances compare the geometry of the unconstrained steady-state flux space and conservation-law space directly, before selecting any particular optimal flux state. The examples below serve as controlled demonstrations that nullspace geometry can contain information that is inaccessible to metrics on metabolic process sets and flux vectors. 

\subsection{One-nutrient chemoton}

Here we consider a simplified form of the chemoton model~\cite{Ganti1975,Csendes1984} with a single nutrient $X$, waste product $Y$, homopolymeric information template $pV_n$, and homopolymeric structural membrane $T_m$. This chemoton consists of a catalytic chemical reaction,
\begin{equation*}
    X \to A + V' + T' + Y ,
\end{equation*}
template synthesis,
\begin{equation*}
    n \, V' \to pV_n + n \, R ,
\end{equation*}
the activation of a structural membrane monomer,
\begin{gather*}
    T' \to T^* \ \ \ \text{and} \ \ \ 
    T^* + R \to T,
\end{gather*}
and structural membrane synthesis,
\begin{equation*}
    m \, T \to T_m .
\end{equation*}
For simplicity, we collapse the cycles of the original model into three single reactions. We allow for the exchange of nutrient source $X$ and waste product $Y$:
\begin{equation*}
    \varnothing \to X \ \ \ \text{and} \ \ \ Y \to \varnothing .
\end{equation*}
We also define the biomass to consist of metabolite $A$ and a monomeric fraction each of the template and structural membrane:
\begin{equation*}
    A + \frac{1}{n} \, pV_n + \frac{1}{m} \, T_m  \to \varnothing .
\end{equation*}
The corresponding stoichiometric matrix is 
\begin{equation*}
S_{n,m} =
\begin{array}{c}
X \\ Y \\ A \\ V' \\ T' \\ pV_n \\ R \\ T^* \\ T \\ T_m 
\end{array}\left[
    \begin{array}{c c c c c  c c c}
        -1 &  0 &  0 &  0 &  0 &  1 &  0 &  0   \\
         1 &  0 &  0 &  0 &  0 &  0 & -1 &  0   \\ 
         1 &  0 &  0 &  0 &  0 &  0 &  0 & -1   \\
         1 & -n &  0 &  0 &  0 &  0 &  0 &  0   \\
         1 &  0 & -1 &  0 &  0 &  0 &  0 &  0   \\
         0 &  1 &  0 &  0 &  0 &  0 &  0 & -n^{-1} \\
         0 &  n &  0 & -1 &  0 &  0 &  0 &  0   \\
         0 &  0 &  1 & -1 &  0 &  0 &  0 &  0   \\
         0 &  0 &  0 &  1 & -m &  0 &  0 &  0   \\
         0 &  0 &  0 &  0 &  1 &  0 &  0 & -m^{-1} 
    \end{array} \right].
\end{equation*}

The right nullspace of $S_{n,m}$ is 
\begin{equation*}
    \mathcal{N}^\text{R}(S_{n,m}) = \mathspan \begin{bmatrix}
        1 \\ n^{-1} \\ 1 \\ 1 \\ m^{-1} \\ 1 \\ 1 \\ 1
    \end{bmatrix}
\end{equation*}
It follows that the Grassmann distance of two such chemoton with different $n,m$ is 
\begin{align*}
&\dist(\mathcal{N}^\text{R}(S_{n_1,m_1}), \mathcal{N}^\text{R}(S_{n_2,m_2})) \\
& \ \ \ \ \ \ = \cos^{-1} \frac{6 n_1m_1n_2m_2 + n_1n_2 + m_1m_2}{\sqrt{6 n_1^2 m_1^2 + n_1^2 + m_1^2}\sqrt{6 n_2^2 m_2^2 + n_2^2 + m_2^2}}.
\end{align*}
In this case, the template condensation and membrane duplication reactions are considered biologically equivalent, despite the fact that the template and structural membrane are different in size. Here, the Jaccard distance would fail to detect a difference. If these reactions are not considered biologically equivalent, the right nullspaces of each chemoton would be augmented to the shared ambient space as: 
\begin{align*}
    \mathcal{N}^\text{R}(S_{n_1,m_1}') &= \mathspan \begin{bmatrix}
        1 \\ n_1^{-1} \\ 0 \\ 1 \\ 1 \\ m_1^{-1} \\ 0  \\ 1 \\ 1 \\ 1
    \end{bmatrix} 
\end{align*}
and 
\begin{align*}
    \mathcal{N}^\text{R}(S_{n_2,m_2}') &= \mathspan \begin{bmatrix}
        1 \\ 0 \\ n_2^{-1} \\ 1 \\ 1 \\ 0 \\ m_2^{-1} \\ 1 \\ 1 \\ 1
    \end{bmatrix}.
\end{align*}
Consequently, the Grassmann distance becomes
\begin{align*}
&\dist(\mathcal{N}^\text{R}(S_{n_1,m_1}'), \mathcal{N}^\text{R}(S_{n_2,m_2}')) \\
& \ \ \ \ \ \ = \cos^{-1} \frac{6 n_1m_1n_2m_2}{\sqrt{6 n_1^2 m_1^2 + n_1^2 + m_1^2}\sqrt{6 n_2^2 m_2^2 + n_2^2 + m_2^2}},
\end{align*}
where we explicitly require $n_1\not=n_2$ and $m_1\not=m_2$. In this scenario, the Jaccard distance ensures that all chemoton models with $n_1 \not= n_2$ and $m_1 \not= m_2$ would be equidistant ($\dist_\text{Jaccard} = 0.4$). Thus the Jaccard distance would be unable to detect differences in the relevant chemoton size scales which are detected by the Grassmann distance due to its inherent sensitivity to stoichiometry.   

\subsection{Two-nutrient chemoton} 
Now consider the related chemoton with two distinct nutrients and corresponding catalytic chemical reactions: waste products: 

\begin{gather*}
       X_1 \to A + V' + T' + Y_1, \\ 
       X_2 \to A + V' + T' + Y_2, \\ 
    n \, V' \to pV_n + n \, R, \\ 
    T' \to T^*,  \\ 
    T^* + R \to T ,\\ 
    m \, T \to T_m ,\\
    \varnothing \to X_1 \ \ \ \text{and} \ \ \  Y_1 \to \varnothing ,\\  
    \varnothing \to X_2 \ \ \ \text{and} \ \ \  Y_2 \to \varnothing, \\ 
    A + \frac{1}{n} \, pV_n + \frac{1}{m} \, T_m  \to \varnothing .
\end{gather*}
The stoichiometric matrix for this reaction network is
\begin{widetext}
\begin{equation*}
S_{n,m}^{(2)} =
\begin{array}{c}
X_1 \\ X_2  \\ Y_1 \\ Y_2 \\ A \\ V' \\ T' \\ pV_n \\ R \\ T^* \\ T \\ T_m 
\end{array}
\left[
    \begin{array}{c c c c c c c c c c c}
        -1 &  0 &  0 &  0 &  0 &  0 &  1 &  0 &  0 &  0 & 0    \\
         0 & -1 &  0 &  0 &  0 &  0 &  0 &  0 &  1 &  0 & 0    \\ 
         1 &  0 &  0 &  0 &  0 &  0 &  0 & -1 &  0 &  0 & 0    \\ 
         0 &  1 &  0 &  0 &  0 &  0 &  0 &  0 &  0 & -1 & 0    \\
         1 &  1 &  0 &  0 &  0 &  0 &  0 &  0 &  0 &  0 & -1   \\
         1 &  1 & -n &  0 &  0 &  0 &  0 &  0 &  0 &  0 & 0    \\
         1 &  1 &  0 & -1 &  0 &  0 &  0 &  0 &  0 &  0 &  0   \\
         0 &  0 &  1 &  0 &  0 &  0 &  0 &  0 &  0 &  0 & -n^{-1} \\
         0 &  0 &  n &  0 & -1 &  0 &  0 &  0 &  0 &  0 &  0   \\
         0 &  0 &  0 &  1 & -1 &  0 &  0 &  0 &  0 &  0 &  0   \\
         0 &  0 &  0 &  0 &  1 & -m &  0 &  0 &  0 &  0 & 0    \\
         0 &  0 &  0 &  0 &  0 &  1 &  0 &  0 &  0 &  0 & -m^{-1} 
    \end{array} \right]
\end{equation*}
\end{widetext}
with two dimensional right nullspace 
\begin{equation}
    \mathcal{N}^\text{R}[S_{n,m}^{(2)}] = \mathspan \left\{ \begin{bmatrix}
        1 \\ 0 \\ n^{-1} \\ 1 \\ 1 \\ m^{-1} \\ 1 \\ 1 \\ 0 \\ 0 \\ 1 
    \end{bmatrix},  
    \begin{bmatrix}
        0 \\ 1 \\ n^{-1} \\ 1 \\ 1 \\ m^{-1} \\ 0 \\ 0 \\ 1 \\ 1 \\ 1
    \end{bmatrix}\right\}.
    \label{eq:2CHEMOTON-RNULL}
\end{equation}

Flux balance analysis seeks a representative element $v_* \in \mathcal{N}^\text{R}[S_{n,m}^{(2)}]$ subject to
\begin{equation*}
    v_* = \argmax_{ v \in \mathcal{N}^\text{R}[S_{n,m}^{(2)}]} f(v) \ \ \ \text{where} \ \ \  f(v) = e_{11}^\intercal v 
\end{equation*}
is an objective function, and $e_{11} = [ 0 , 0 , 0 , \ldots, 1]^\intercal$ is the 11-dimensional standard basis vector that selects the flux through the biomass process. As $v_*$ is an element of $\mathcal{N}^\text{R}[S_{n,m}^{(2)}]$, it can be written as a linear combination of linearly independent elements $v_1, v_2$: 
\begin{equation*}
    v = a_1 v_1 + a_2 v_2 .
\end{equation*}
Without loss of generality, we take $v_1,v_2$ as in Eq.~(\ref{eq:2CHEMOTON-RNULL}). It follows that 
\begin{equation*}
    f = e_{11}^\intercal(a_1 v_1 + a_2 v_2 ) = a_1 + a_2 .
\end{equation*}

Now, let us define the nutrient influxes in the following manner: 
\begin{equation*}
    e_7^\intercal v = a_1  = 1 \ \ \ \text{and} \ \ \ e_9^\intercal v = a_2 = \rho ,
\end{equation*}
where $\rho \in \mathbb{Z}_{>0}$ tunes the strength of the influx of $X_2$ relative to $X_1$. Under these definitions, the objective function simplifies to
$$f = 1 + \rho.$$ 
Thus, the solution to the FBA problem is 
\begin{equation*} v_* = 
    \begin{bmatrix}
        1 \\ \rho  \\ (1+\rho )n^{-1} \\ 1 + \rho  \\ 1 + \rho  \\ (1+\rho )m^{-1} \\ 1 \\ 1 \\ \rho  \\ \rho  \\ 1 + \rho  
    \end{bmatrix} .
\end{equation*}

Setting $\rho \to 0$, we recover the FBA solution for the one-nutrient chemoton 
\begin{equation*}
    w_* = (1+\rho) \begin{bmatrix}
        1 \\ 0 \\ n^{-1} \\ 1 \\ 1 \\ m^{-1} \\ 1 \\ 1 \\ 0 \\ 0 \\ 1 
    \end{bmatrix},
\end{equation*}
augmented to be consistent with the flux space of $v_*$ and scaled by $1+\rho$ so that the total influxes in both one- and two-nutrient chemotons are equal. We also assume that the two chemotons have identically sized templates and membranes, for simplicity. 

If the absolute flux values are relevant, the natural choice for distance between the FBA solutions is the $\ell^2$-norm: 
\begin{equation*}
    \dist_{\ell^2}(v_*,w_*) = \lVert v_* - w_* \rVert_2
    = \sqrt{6} \rho .
\end{equation*}
Otherwise, comparing the relative distribution of fluxes through the different chemotons suggests selecting the angular distance:
\begin{align*}
    &\dist_\text{angular}(v_*,w_*) = \cos^{-1} \frac{v_*^\intercal w_*}{\lVert v_* \rVert_2 \lVert w_* \rVert_2} \\ & \ \ \ =
     \cos^{-1} 
    \frac{(1+\rho)(n^2 + m^2) + 3 (2+\rho) n^2 m^2 }{n^2m^2
    \sqrt{(6 + \varsigma)[6 (1+\rho + \rho^2) + (1+\rho)^2 \varsigma ] }
    },
\end{align*}
where $\varsigma = n^{-2} + m^{-2}$. The angular distance between FBA solutions is in the spirit of the R-Grassmann distance where instead of comparing two nullspaces, two representative elements are selected and compared. 

In either case, the distances between FBA solutions explicitly depend on the value of the nutrient influxes. This makes distances between FBA solutions powerful when information about nutrient abundance is available, but a singular solution may not be identifiable without this information. Moreover, by construction, FBA requires that solutions are optimal for a pre-defined objective function. Consequently, FBA solutions may depend on the choice of objective function; then which objective function is the ``correct'' one? We may have instead optimized for the efflux of waste product $Y_2$ in the two-nutrient chemoton, as in the bioengineered scenario where $Y_2$ is a desirable product of metabolism. The resulting FBA solution would be orthogonal to the FBA solution for the one-nutrient chemoton with the optimization performed for biomass production. This would in turn result in a constant but nonzero angular distance between the FBA solutions ($\dist_\text{angular} = \pi/2$). In this case, the angular distance is equivalent to the R-Grassmann distance which is the direct result of the unity nullspace dimension difference and the subsumed nullspaces ($\mathcal{N}^{\text{R}}[S_{n,m}^{(1)}] \subseteq \mathcal{N}^\text{R}[S_{n,m}^{(2)}]$). This comparison emphasizes that distances over FBA solutions and right nullspaces answer different questions. The former compare context- and objective-dependent metabolic states, whereas the latter compare the unconstrained steady-state flux spaces from which such states are drawn. Thus, flux-based and Grassmann-based comparisons are complementary: flux-based metrics are preferable when a specific environment and objective are known, whereas the R-Grassmann distance provides an objective-independent comparison of the unconstrained steady-state flux space.

\subsection{Two-nutrient two-metabolite chemoton}

Let us now modify the two-nutrient chemoton such that each catalytic chemical reaction produces a different metabolite that is required in the biomass:
\begin{gather*}
       X_1 \to A_1 + V' + T' + Y_1, \\ 
       X_2 \to A_2 + V' + T' + Y_2, \\ 
    n \, V' \to pV_n + n \, R, \\ 
    T' \to T^*,  \\ 
    T^* + R \to T ,\\ 
    m \, T \to T_m ,\\
    \varnothing \to X_1 \ \ \ \text{and} \ \ \  Y_1 \to \varnothing ,\\  
    \varnothing \to X_2 \ \ \ \text{and} \ \ \  Y_2 \to \varnothing, \\ 
    A_1 + \lambda A_2 + \frac{1+\lambda}{n} \, pV_n + \frac{1+\lambda}{m} \, T_m  \to \varnothing .
\end{gather*}
Here $\lambda \in \mathbb{Z}_{>0}$ is the relative demand for the two metabolites $A_1$ and $A_2$. 

The left nullspace of the corresponding matrix $S_{n,m}^{'(2)}$ is 
\begin{equation*}
    \mathcal{N}^\text{L}[S_{n,m}^{'(2)}] = \mathspan
    \left\{ 
    \begin{bmatrix}
        0 \\ 0 \\ 0 \\ 0 \\ -1 \\ -1 \\ 1 \\ 0 \\ n \\ 0 \\ 0 \\ 0 \\ 0 
    \end{bmatrix}, 
    \begin{bmatrix}
        0 \\ 0 \\ 0 \\ 0 \\ -1 \\ -1 \\ 1 \\ 0 \\ 0 \\ 1 \\ 0 \\ 1 \\ m
    \end{bmatrix}, 
    \begin{bmatrix}
        0 \\ 0 \\ 0 \\ 0 \\ 0 \\ 0 \\ -1 \\ 1 \\ 0 \\ -1 \\ 1 \\ 0 \\ 0 
    \end{bmatrix} \right\} ,
\end{equation*}
which is contrasted by the left nullspace of the two-nutrient one-metabolite chemoton
\begin{equation*}
    \mathcal{N}^\text{L}[S_{n,m}^{'(2)}] = \mathspan
    \left\{ 
    \begin{bmatrix}
        0 \\ 0 \\ 0 \\ 0 \\ -1 \\ 0 \\ 1 \\ 0 \\ n \\ 0 \\ 0 \\ 0 \\ 0 
    \end{bmatrix}, 
    \begin{bmatrix}
        0 \\ 0 \\ 0 \\ 0 \\ -1 \\ 0 \\ 1 \\ 0 \\ 0 \\ 1 \\ 0 \\ 1 \\ m
    \end{bmatrix}, 
    \begin{bmatrix}
        0 \\ 0 \\ 0 \\ 0 \\ 0 \\ 0 \\ -1 \\ 1 \\ 0 \\ -1 \\ 1 \\ 0 \\ 0 
    \end{bmatrix} \right\} ,
\end{equation*}
augmented to be consistent with the conservation space of $S_{n,m}^{'(2)}$.  

For simplicity, let us consider two regimes: (1) $m\to\infty$ and (2) $n\to\infty$. In the case where the structural membrane is infinitely large, we find that the L-Grassmann distance between the two-nutrient chemotons is
\begin{equation*}
    \dist(\mathcal{N}^\text{L}[S_{n,\infty}^{(2)}], \mathcal{N}^\text{L}[S_{n,\infty}^{'(2)}]) = \cos^{-1}\sqrt{\frac{7 + 4n ^2}{11 + 4 n^2}},
\end{equation*}
whereas when the template is infinitely large the distance is 
\begin{equation*}
    \dist(\mathcal{N}^\text{L}[S_{\infty,m}^{(2)}], \mathcal{N}^\text{L}[S_{\infty,m}^{'(2)}]) = \cos^{-1}\sqrt{\frac{3 + m^2}{4 + m^2}}.
\end{equation*}
When both $n,m\to\infty$, the two conservation spaces are identical, $\dist(\mathcal{N}^\text{L}[S_{\infty,\infty}^{(2)}], \mathcal{N}^\text{L}[S_{\infty,\infty}^{'(2)}]) = 0$. In either case, the distances are not a function of the relative demand $\lambda$ for the two metabolites in the biomass. Similarly, the Jaccard distance is unable to capture this dependence as it is not sensitive to stoichiometries. Instead, we turn to the R-Grassmann distance which has the same form in both regimes:
\begin{align*}
     \dist(\mathcal{N}^\text{R}[S_{\nu,\infty}^{(2)}], \mathcal{N}^\text{R}[S_{\nu,\infty}^{'(2)}])
     &= \dist(\mathcal{N}^\text{R}[S_{\infty,\nu}^{(2)}], \mathcal{N}^\text{R}[S_{\infty,\nu}^{'(2)}]) \\
     &= \sqrt{\frac{\pi^2}{4} + \vartheta(\nu,\lambda)^2}
\end{align*}
where 
\begin{widetext}
\begin{equation*}
    \vartheta(\nu,\lambda) = \cos^{-1} \sqrt{\frac{ 2(1+\lambda)^2 + [21 + \lambda(26 + 16 \lambda)]\nu^2 + 2[27 + \lambda(18 + 19 \lambda)]\nu^4 }{(2 + 9 \nu^2)[(1+\lambda)^2 + (6 + \lambda (4+5\lambda))\nu^2]}} ,
\end{equation*}
\end{widetext}
is the principal angle between the right nullspaces. Unlike the left nullspace, the flux spaces are never identical, even in the doubly-infinite regime, where the principal angle becomes 
\begin{equation*}
     \vartheta(\infty,\lambda) =  \cos^{-1} \sqrt{\frac{54 + 36 \lambda + 38\lambda^2}{54 + 36 \lambda + 45 \lambda^2}},
\end{equation*}
because of the difference in nullity. Thus, the L-Grassmann and R-Grassmann distances can provide distinct and complementary information about differences between metabolic networks. These analytically solvable examples also demonstrate that nullspace geometry can distinguish network differences that are invisible, degenerate, or difficult to interpret with metabolic process-overlap comparisons alone.

\begin{figure}
\centering
\includegraphics[width=8.7cm]{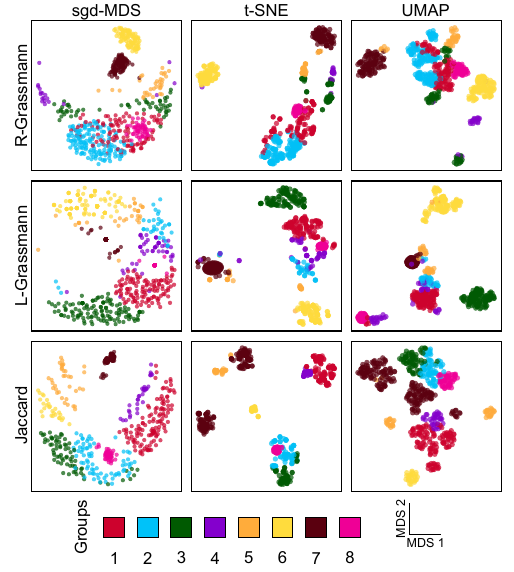}
    \caption{\textbf{Embeddings of distances highlight that metabolic groups are robust to the choice of embedding procedure.} Colors correspond to metabolic groups identifies in Fig.~\ref{fig:metabolic-clusterings}. t-SNE complexity and learning rate are both set to 50. UMAP \# neighbors and minimum distance are set to 10 and 2, respectively. 
    }
    \label{fig:tSNE-UMAP}
\end{figure}
\begin{figure*}[bt!]
\centering
\includegraphics[width=17.8cm]{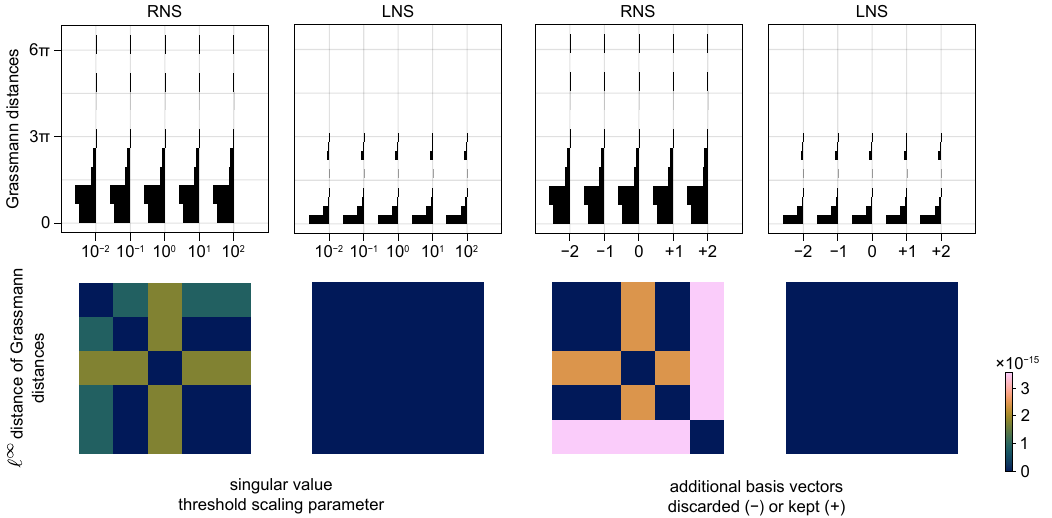}
    \caption{\textbf{Grassmann distances are numerically stable.} Nullspaces are numerically computed via rank-thresholding for 267 unique \textit{in silico} KO metabolic networks with alterations to the singular value threshold (left) or cutoff basis vector (right) around the default values (Appendix~\ref{sec:num-stability}). The default singular value threshold for the rank is determined from the size of the stoichiometric matrix $S \in \mathbb{R}^{m\times n}$: $\epsilon \min(m,n)$ where $\epsilon \approx 2.22\times 10^{-16}$ is the machine epsilon of the \textsc{Julia} \texttt{Float64} type. Note that changes in the singular value threshold (left) can lead to heterogeneous changes in the number of basis vectors, which is not the case when additional vectors are discarded or kept (right). In the latter, the dimension gap remains strictly unchanged. Distributions of the Grassmann distances (top) appear equal under these alterations with a maximum difference in distances on the order of $10^{-15}$ (bottom). Abbr: RNS = right nullspace, LNS = left nullspace. }
    \label{fig:sensitivity}
\end{figure*}

\section{STOCHASTIC GRADIENT DESCENT MULTIDIMENSIONAL SCALING\label{sec:sgd-MDS}}

We embed distance matrices by stochastic gradient descent multidimensional scaling (sgd-MDS) with loss function 
\begin{equation*}\textstyle 
    \mathcal{L}(u) = 2 \sum_{\langle i, j \rangle} w_{ij} (\dist_\mathcal{M}(u_i,u_j) - d_{ij} )^2
\end{equation*}
where $\dist_{\mathcal{M}}$ is the distance function on the manifold $\mathcal{M}$ and $w_{ij} = d_{ij}^{-2}$ are the weights for each pair of points~\cite{Torgerson1952}. To avoid numerical s, weights are adjusted $w_{ij} = (d_{ij} + \epsilon)^2$, where $\epsilon \approx 2.22 \times 10^{-16}$ is the machine epsilon of the \textsc{Julia} \texttt{Float64} type. For Euclidean manifolds $\mathcal{M} = \mathbb{R}^n$, the distance function is the $\ell^2$-norm. Taking a physical prescription, this loss function is equivalent to the Hamiltonian of a system of fully coupled springs with spring constants $w_{ij}$ and equilibrium lengths $d_{ij}$. The optimization process is then a relaxation of the system to a minimum of the Hamiltonian as described in Ref.~\cite{Zheng2019}. This is achieved by iterative gradient descent steps for each spring in a random permutation---with replacement---according to an exponentially decay annealing schedule $\eta(t) = e^{-\lambda t}$ of step sizes $\mu_{ij} = \eta w_{ij}$ for a fixed number of steps $t = 0 \to t_{\text{s}} - 1$ where $t_{\text{s}} =10^3$ for mini-AGORA2 networks, $t_{\text{s}} =10^5$ for organ networks, and $t_{\text{s}} =10^6$ for planetary atmosphere networks. The exponential decay rate is chosen to be $\lambda = (t_{\text{s}}-1)^{-1} \log (\max w_{ij}/\varepsilon \min w_{ij})$, where $\varepsilon = 0.1$. To allow for convergence, this is followed by an annealing schedule of $\Theta(1/t)$ until the relative change in the loss function drops below a threshold $\delta = 10^{-8}$ or a fixed number of iterations $\tau_{\text{s}}=t_{\text{s}}$ are reached. We note that the spring relaxation approach described in Ref.~\cite{Zheng2019} can be generalized by confining the motions to manifolds with well-defined exp and log maps: 
\begin{gather*}
    r_{ab} \leftarrow \left[ \left\{ 1- \frac{1}{2} \mu \right\} \dist_\mathcal{M}(u_a,u_b) + \frac{1}{2} \mu d_{ab} \right] \frac{\log_{u_a} u_b}{\dist_\mathcal{M}(u_a,u_b)} \\
    u_i \leftarrow \exp_{u_j} r_{ji} \\
    u_j \leftarrow \exp_{u_i} r_{ij}
\end{gather*}

We compare our multidimensional scaling algorithm against t-SNE and UMAP for embedding bacterial metabolic distances and find that cluster placement is consistent across embeddings [Fig.~\ref{fig:tSNE-UMAP}]. Importantly, all clusters analyzed in the main text are computed from distance matrices rather than embeddings; the embeddings are used only for visualization.

\section{NUMERICAL CONSIDERATIONS OF THE GRASSMANN DISTANCE \label{sec:num-stability}}

To compute the Grassmann distance between metabolic networks, we first compute stoichiometric nullspaces. Computing a right nullspace orthonormal basis of a matrix $M \in \mathbb{R}^{m\times n}$ requires a singular value decomposition (SVD):
\begin{equation*}
    M = U \Sigma V^\intercal,
\end{equation*}
where $U \in \mathbb{R}^{m\times m}$ and $V\in \mathbb{R}^{n\times n}$ are orthogonal matrices, and $\Sigma \in \mathbb{R}^{m\times n}$ is a rectangular diagonal matrix of singular values. Let $\sigma \in \mathbb{R}^{\min(m,n)} = \diag(\Sigma)$ be the diagonal entries of $\Sigma$. By convention $\sigma_1 \geq \sigma_2 \geq \ldots \geq \sigma_{\min(m,n)}$. 

We then identify the rank of $M$ which is equal to the number of nonzero singular values and requires defining a numerical zero for thresholding. Rescaling by the largest singular value, $\tilde{\sigma} = [1, \sigma_2/\sigma_1,\ldots, \sigma_{\min(m,n)}/\sigma_1]^\intercal$, we threshold the singular values by $\epsilon \min(m,n)$, where $\epsilon$ is the machine epsilon of the \textsc{Julia} \texttt{Float64} type. The rank of $M$ is then
\begin{equation*}
    \rank(M) = \sum_{i = 1}^{\min(m,n)} \mathbbm{1}[ \tilde{\sigma}_i > \epsilon \min(m,n)] ,
\end{equation*}
where $\mathbbm{1}$ is the indicator function. The last $n-\rank(M)$ columns of $V$ are the desired right nullspace basis vectors. Simultaneously, we can compute the left nullspace basis vectors which are the last $m-\rank(M)$ columns of $U$. In this manner, we compute rank-thresholded nullspaces.  

To demonstrate that the Grassmann distance is numerically stable, we alter the choice of numerical zero in thresholding by introducing a scaling parameter, $\epsilon \min(m,n) \to a \epsilon \min(m,n)$, and allow $a$ to vary in magnitude from $10^{-2} \to 10^{2}$. In this range, we find no appreciable difference between the distances calculated for the 267 unique $\textit{in silico}$ KO metabolic networks of \textit{Escherichia coli} K-12 MG1655 [Fig.~\ref{fig:sensitivity}, left]. Similarly, if we directly perturb the rank of $M$ and keep or discard additional vectors, only the angular term changes because the dimension gap remains the same and we find no appreciable difference in the distances [Fig.~\ref{fig:sensitivity}, right]. Here, no left nullspaces are trivial, so we need not account for this case in the removal of basis vectors. 

\section{THE SQUARE ROOT OF A METRIC IS A METRIC\label{sec:square-root-metric}}

We take the square root of the tree graph metric as a definition of phylogenetic distance. We prove here that this distance is a metric by showing that, more generally, the square root ($\sqrt{} \, : \, \mathbb{R}_{\geq 0} \to \mathbb{R}_{\geq 0}$) of any metric is a metric. Suppose that we have a metric $d$ that, by definition, satisfies the axioms: (1) $d(x,x) = 0$, (2) $x\not= y \implies d(x,y) > 0$, (3) $d(x,y) = d(y,x)$, and (4) $d(x,z) \leq d(x,y) + d(y,z)$. Because the square root is nonnegative, symmetric to swapping arguments, and vanishes only at zero, the only nontrivial metric property is the triangle inequality. For any metric $\sqrt{d}$,

\begin{align*}
    \sqrt{d(x,z)} 
    &\leq \sqrt{d(x,y) + d(y,z)} \\
    &\leq \sqrt{d(x,y) + d(y,z) + 2\sqrt{d(x,y)d(y,z)}}  \\
    &= \sqrt{\left(\sqrt{d(x,y)} + \sqrt{d(y,z)}\right)^2} \\
    &= \sqrt{d(x,y)} + \sqrt{d(y,z)} .
\end{align*}

Thus, $\sqrt{d}$ is also metric.\vfill

\section{CLOSED-NETWORK LEFT NULLSPACE ANALYSIS AS AN AUXILIARY PROBE OF LATENT CONSERVATION STRUCTURE \label{sec:open-network-groups}}

Here we consider the unique L-Grassmann groups of closed bacterial networks identified in Fig.~\ref{fig:open-vs-closed}(c). These results are an auxiliary analysis of latent internal conservation structure after removing exchange processes. Group 2 is marked by a depletion of 3-oxoacyl-ACP synthase and reductase (n-C14:0). Enrichment of 3-oxoacyl-ACP synthase and 3-oxoacyl-ACP reductase consistent with increased representation of bacterial type II fatty-acid biosynthesis, a pathway required for membrane lipid production and cell-envelope maintenance~\cite{Parsons2013}. \vfill

% Bibliography

%

\end{document}